%% file: oext.tex
\documentclass[11pt]{article}
\usepackage{amsmath,amsthm,bm, amscd}
\usepackage{fullpage}
\usepackage[nofullpage,full,titlepage,nousetoc,nouselot,nouselof,hylinks,final]{boaz}

\usepackage{color}

\usepackage{graphicx}

\newcommand{\conv}{\mathsf{Conv}}
\newcommand{\supp}{\mathsf{supp}}

\newcommand{\twext}{\mathsf{2Ext}}

\newcommand{\iext}{\mathsf{IExt}}

\newcommand{\zo}{\bits}

\newcommand{\acb}{\mathsf{AdvCB}}
\newcommand{\adg}{\mathsf{AdvGen}}
\newcommand{\RS}{\mathsf{RS}}

\newcommand{\snmExt}{\textnormal{snmExt}}
\newcommand{\nipm}{\textnormal{NIPM}}
\newcommand{\A}{\mathcal{A}}
\newcommand{\scirc}{\hspace{0.1cm}\circ \hspace{0.1cm}}
\newcommand{\X}{\mathbf{X}}
\newcommand{\rr}{\mathbf{R}}
\newcommand{\Y}{\mathbf{Y}}

\newcommand{\W}{\mathbf{W}}
\newcommand{\snm}{\mathsf{NM}}
\newcommand{\flip}{\textnormal{flip-flop}}
\newcommand{\same}{\textnormal{$same^{\star}$}}
\newcommand{\cpy}{\textnormal{copy}}

\def\calX{{\mathcal X}}
\def\calY{{\mathcal Y}}

\DeclareMathOperator{\nm}{\nmExt} 
\DeclareMathOperator{\expect}{E}

\newcommand{\laext}{\mathsf{laExt}}
\newcommand{\samp}{\mathsf{Samp}}

\theoremstyle{definition}

\newcommand{\eps}{\epsilon}

\newcommand{\Cond}{\mathsf{Cond}}

\newcommand{\Supp}{\mathsf{Supp}}
\newcommand{\raz}{\mathsf{Raz}}

\newcommand{\Ext}{\mathsf{Ext}}

\newcommand{\nmExt}{\mathsf{nmExt}}
\newcommand{\mac}{\mathsf{MAC}}
\newcommand{\bip}{\mathsf{IP}}
\newcommand{\Enc}{\mathsf{Enc}}
\newcommand{\Dec}{\mathsf{Dec}}

\newcommand{\TExt}{\mathsf{TExt}}
\newcommand{\zuc}{\Cond}
\newcommand{\scond}{\mathsf {Scond}}

\newcommand{\BI}{\begin{itemize}}
\newcommand{\EI}{\end{itemize}}
\newcommand{\BE}{\begin{enumerate}}
\newcommand{\EE}{\end{enumerate}}

\newtheorem{thm}{Theorem}      
\newcommand{\BT}{\begin{theorem}}   \newcommand{\ET}{\end{theorem}}
\newcommand{\BD}{\begin{definition}}   \newcommand{\ED}{\end{definition}}
\newcommand{\BCR}{\begin{corollary}} \newcommand{\ECR}{\end{corollary}}
\newtheorem{constr}[thm]{Construction}
\newcommand{\BCT}{\begin{constr}} \newcommand{\ECT}{\end{constr}}
\newcommand{\BL}{\begin{lemma}}   \newcommand{\EL}{\end{lemma}}

\newcommand{\BP}{\begin{proposition}}   \newcommand{\EP}{\end{proposition}}
\newcommand{\BCM}{\begin{claim}}   \newcommand{\ECM}{\end{claim}}
\newcommand{\BF}{\begin{fact}}   \newcommand{\EF}{\end{fact}}
\newcommand{\BA}{\begin{assumption}}   \newcommand{\EA}{\end{assumption}}

\def\eps{\varepsilon}

\def\le{\leqslant} \def\ge{\geqslant}

\makeatletter
\def\ExtendSymbol#1#2#3#4#5{\ext@arrow 0099{\arrowfill@#1#2#3}{#4}{#5}}
\def\RightExtendSymbol#1#2#3#4#5{\ext@arrow 0359{\arrowfill@#1#2#3}{#4}{#5}}
\def\LeftExtendSymbol#1#2#3#4#5{\ext@arrow 6095{\arrowfill@#1#2#3}{#4}{#5}}
\makeatother
\newcommand\llrightarrow[2][]{\RightExtendSymbol{-}{-}{\rightarrow}{#1}{#2}}

\newcommand\llleftarrow[2][]{\RightExtendSymbol{\leftarrow}{-}{-}{#1}{#2}}

\newcommand{\hinf}{H_\infty}
\newcommand{\thinf}{\widetilde{H}_\infty}

\begin{document}

\begin{titlepage}
\def\thepage{}

\date{}
\title{Non-Malleable Extractors and Non-Malleable Codes: Partially Optimal Constructions}

\author{
Xin Li \thanks{Supported by NSF award CCF-1617713.}\\
Department of Computer Science\\
Johns Hopkins University\\
Baltimore, MD 21218, U.S.A.\\
lixints@cs.jhu.edu
}

\maketitle \thispagestyle{empty}

\begin{abstract}
The recent line of study on randomness extractors has been a great success, resulting in exciting new techniques, new connections, and breakthroughs to long standing open problems in several seemingly different topics.\ These include seeded non-malleable extractors, privacy amplification protocols with an active adversary, independent source extractors (and explicit Ramsey graphs), and non-malleable codes in the split state model.  Previously, the best constructions are given in \cite{Li17}: seeded non-malleable extractors with seed length and entropy requirement $O(\log n+\log(1/\e)\log \log (1/\e))$ for error $\e$; two-round privacy amplification protocols with optimal entropy loss for security parameter up to $\Omega(k/\log k)$, where $k$ is the entropy of the shared weak source; two-source extractors for entropy $O(\log n \log \log n)$; and non-malleable codes in the $2$-split state model with rate $\Omega(1/\log n)$. However, in all cases there is still a gap to optimum and the motivation to close this gap remains strong. 

In this paper, we introduce a set of new techniques to further push the frontier in the above questions. Our techniques lead to improvements in all of the above questions, and in several cases partially optimal constructions. This is in contrast to all previous work, which only obtain close to optimal constructions. Specifically, we obtain:  
\begin{enumerate}
\item A seeded non-malleable extractor with seed length $O(\log n)+\log^{1+o(1)}(1/\e)$ and entropy requirement $O(\log \log n+\log(1/\e))$, where the entropy requirement is asymptotically optimal by a recent result of Gur and Shinkar \cite{GurS17}; \item A two-round privacy amplification protocol with optimal entropy loss for security parameter up to $\Omega(k)$, which solves the privacy amplification problem completely;\footnote{Except for the communication complexity, which is of secondary concern to this problem.} \item A two-source extractor for entropy $O(\frac{\log n \log \log n}{\log \log \log n})$, which also gives an explicit Ramsey graph on $N$ vertices with no clique or independent set of size $(\log N)^{O(\frac{\log \log \log N}{\log \log \log \log N})}$; and \item The first explicit non-malleable code in the $2$-split state model with \emph{constant} rate, which has been a major goal in the study of non-malleable codes for quite some time. One small caveat is that the error of this code is only (an arbitrarily small) constant, but we can also achieve negligible error with rate $\Omega(\log \log \log n/\log \log n)$, which already improves the rate in \cite{Li17} exponentially. 
\end{enumerate}
We believe our new techniques can help to eventually obtain completely optimal constructions in the above questions, and may have applications in other settings. 

\end{abstract}
\end{titlepage}

\section{Introduction}
The study of randomness extractors has been a central line of research in the area of pseudorandomness, where the goal is to understand how to use randomness more efficiently in computation. As fundamental objects in this area, randomness extractors are functions that transform imperfect random sources into nearly uniform random bits.\ Their original motivation is to bridge the gap between the uniform random bits required in standard applications (such as in randomized algorithms, distributed computing, and cryptography), and practical random sources which are almost always biased (either because of natural noise or adversarial information leakage). However the study of these objects has led to applications far beyond this motivation, in several different fields of computer science and combinatorics (e.g., coding theory, graph theory, and complexity theory).

The inputs to a randomness extractor are usually imperfect randomness, modeled by the notion of general weak random sources with a certain amount of entropy. 

\begin{definition}
The \emph{min-entropy} of a random variable~$X$ is
\[ H_\infty(X)=\min_{x \in \supp(X)}\log_2(1/\Pr[X=x]).\]
For $X \in \zo^n$, we call $X$ an $(n,H_\infty(X))$-source, and we say $X$ has
\emph{entropy rate} $H_\infty(X)/n$.
\end{definition}

An extensively studied model of randomness extractors is the so called \emph{seeded extractors}, introduced by Nisan and Zuckerman \cite{NisanZ96}. The inputs to a seeded extractor are a general weak random source and a short independent uniform random seed. The random seed is necessary here since it is well known that no deterministic extractor with one general weak source as input can exist. Such extractors have wide applications in computer science. 

\begin{definition}(Seeded Extractor)\label{def:strongext}
A function $\Ext : \bits^n \times \bits^d \rightarrow \bits^m$ is  a \emph{$(k,\eps)$-extractor} if for every source $X$ with min-entropy $k$
and independent $Y$ which is uniform on $\zo^d$,
\[|\Ext(X, Y)-U_m | \leq \e.\]
If in addition we have $|(\Ext(X, Y), Y) - (U_m, Y)| \leq \e$ then we say it is a \emph{strong $(k,\eps)$-extractor}.
\end{definition}

Through a long line of research, we now have explicit constructions of seeded extractors with almost optimal parameters (e.g., \cite{LuRVW03, GuruswamiUV09, DvirW08, DvirKSS09}). In the last decade or so, the focus has shifted to several different but related models of randomness extractors, including seedless extractors and non-malleable extractors. The study of these topics has also been quite fruitful, leading to breakthroughs to several long standing open problems. 

\subsection{Seedless extractors}
As the name suggests, a seedless extractor uses no uniform seed, and the only inputs are weak random sources. Here, again we have two different cases. In the first case, one puts additional restrictions on a single weak random source in order to allow possible extraction, thus obtaining deterministic extractors for special classes of (structured) sources. In the second case, the sources are still general weak random sources, but the extractor needs to use more than one sources. To make extraction possible, one typically assumes the input sources to the extractor are independent, and this kind of extractors are sometimes called independent source extractors. 

Since the pioneering work of Chor and Goldreich \cite{ChorG88}, the study of independent source extractors has gained significant attention due to their close connections to explicit Ramsey graphs, and their applications in distributed computing and cryptography with general weak random sources \cite{KalaiLRZ08, KalaiLR09}.\ The goal here is to give explicit constructions that match the probabilistic bound: an extractor for just two independent $(n, k)$ sources with $k \geq \log n+O(1)$ that outputs $\Omega(k)$ bits with exponentially small (in $k$) error.\ Note that an explicit two-source extractor for such entropy (even with one bit output and constant error) will give an (strongly) explicit Ramsey graph on $N$ vertices with no clique or independent set of size $O(\log N)$, solving an open problem proposed by Erd\H{o}s \cite{erdos:ramsey} in his seminal paper that inaugurated the probabilistic
method.

While early progress on this problem has been quite slow, with the best known construction in almost 20 years only able to handle two independent $(n, k)$ sources with $k > n/2$ \cite{ChorG88}, since 2004 there has been a long line of work \cite{BarakIW04, BarakKSSW05, Raz05, Bourgain05, Rao06, BarakRSW06, Li11b, Li12b, Li13a, Li13b, Li15, Cohen15, CZ15, Li16, Coh16b, CL16, Coh16, BDT16, Cohen16, Li17} introducing exciting new techniques to this problem. This line of work greatly improved the situation and led to a series of breakthroughs. Now we have three source extractors for entropy $k \geq \polylog(n)$ that output $\Omega(k)$ bits with exponentially small error \cite{Li15}, two-source extractors for entropy $k \geq \polylog(n)$ that output $\Omega(k)$ bits with polynomially small error \cite{CZ15, Li16, Mek:resil}, and two-source extractors for entropy $k \geq O(\log n \log \log n)$ that output one bit with any constant error \cite{Li17}. This also gives an explicit Ramsey graph on $N$ vertices with no clique or independent set of size $(\log N)^{O(\log \log \log N)}$. Interestingly and somewhat surprisingly, the most recent progress which brought the entropy requirement close to optimal, has mainly benefited from the study of another kind of extractors, the so called \emph{non-malleable extractors}, which we now describe below. 

\subsection{Non-malleable extractors}
Non-malleable extractors are strengthening of standard extractors, where one requires that  the output is close to uniform even given the output of the extractor on tampered inputs. 

\begin{definition}[Tampering Funtion]
For any function $f:S \rightarrow S$, We say $f$ has no fixed points if $f(s) \neq s$ for all $s \in S$. For any $n>0$, let $\mathcal{F}_n$ denote the set of all functions $f: \{ 0,1\}^n \rightarrow \{0,1\}^n$. Any subset of $\mathcal{F}_n$ is a family of  tampering functions. 
\end{definition}

Depending on what the tampering function acts on, we also have different models of non-malleable extractors. If the tampering acts on the seed of a seeded extractor, such extractors are called \emph{seeded non-malleable extractors}, originally introduced by Dodis and Wichs \cite{DW09}.

\begin{definition} A function $\snmExt:\{0,1\}^n \times \{ 0,1\}^d \rightarrow \{ 0,1\}^m$ is a seeded non-malleable extractor for min-entropy $k$ and error $\epsilon$ if the following holds:\ If $X$ is an $(n, k)$ source and $\A : \{0,1\}^d \rightarrow \{0,1\}^d $ is an arbitrary tampering function with no fixed points, then
$$  \left |\snmExt(X,U_d) \scirc \snmExt(X,\A(U_d))  \scirc U_d- U_m \scirc  \snmExt(X,\A(U_d)) \scirc U_d \right | <\epsilon $$where $U_m$ is independent of $U_d$ and $X$.
\end{definition}

If the tampering acts on the sources of an independent source extractor, then we have \emph{seedless non-malleable extractors}, originally introduced by Cheraghchi and Guruswami \cite{CG14b}.  

\begin{definition}\label{def:t2}
A function $\nmExt : (\{ 0,1\}^{n})^C \rightarrow \{ 0,1\}^m$ is a $(k, \e)$-seedless non-malleable extractor for $C$ independent sources, if it satisfies the following property: Let $X_1, \cdots, X_C$ be $C$ independent $(n, k)$ sources, and $f_1, \cdots, f_C : \zo^n \to \zo^n$ be $C$ arbitrary tampering functions such that there exists an $f_i$ with no fixed points,\footnote{The original definition of seedless non-malleable independent source extractors in \cite{CG14b} allows fixed points, but the two definitions are equivalent up to a small loss in parameters. See Section~\ref{sec:nmtext} for details.} then  $$ |\nmExt(X_1, \cdots, X_C) \circ \nmExt(f_1(X_1), \cdots, f_C(X_2)) - U_m \circ \nmExt(f_1(X_1), \cdots, f_C(X_2))| < \epsilon.$$ 

\end{definition}

\paragraph{Seeded non-malleable extractors and privacy amplification.} Seeded non-malleable extractors were introduced by Dodis and Wichs \cite{DW09}, to study the basic problem of \emph{privacy amplification} \cite{BennettBR88}.  Consider the situation where two parties with local (non-shared) uniform random bits try to convert a shared secret weak random source $\X$ into shared secret uniform random bits. They do this by communicating through a channel, which is watched by an adversary with unlimited computational power. Standard strong seeded extractors provide very efficient protocols for a passive adversary (i.e., can only see the messages but cannot change them), but fail for an active adversary (i.e., can arbitrarily change, delete and reorder messages). In the latter case, which is the focus of this paper, the main goal is to design a protocol that uses as few number of interactions as possible, and achieves a shared uniform random string $\rr$ which has \emph{entropy loss} (the difference between the length of the output and $H_{\infty}(\X)$) as small as possible. Such a protocol is defined with a security parameter $s$, which means the probability that an active adversary can successfully make the two parties output two different strings without being detected is at most $2^{-s}$. On the other hand, if the adversary remains passive, then the two parties should achieve a shared secret string that is $2^{-s}$-close to uniform.\ We refer the reader to \cite{DLWZ11} for a formal definition.

A long line of work has been devoted to this problem \cite{MW97,dkrs,DW09,RW03,KR09,ckor,DLWZ11,CRS11,Li12a,Li12b,Li15b, CGL15, Coh15nm, Coh16a, CL16, Coh16, Cohen16, Li17}. It is known that one round protocol can only exist when the entropy rate of $\X$ is bigger than $1/2$, and the protocol has to incur a large entropy loss. When the entropy rate of $\X$ is smaller than $1/2$, \cite{DW09} showed that any protocol has to take at least two rounds with entropy loss at least $\Omega(s)$. Achieving a two-round protocol with entropy loss $O(s)$ for all possible security parameters $s$ is thus the holy grail of this problem (note that $s$ can be at most $\Omega(k)$ where $k=H_{\infty}(\X)$).

While early works on this problem used various techniques, in \cite{DW09}, Dodis and Wichs introduced a major tool, the seeded non-malleable extractor defined above. They showed that two-round privacy amplification protocols with optimal entropy loss can be constructed using explicit seeded non-malleable extractors. Furthermore, non-malleable extractors exist when $k>2m+2\log(1/\eps) + \log d + 6$ and $d>\log(n-k+1) + 2\log (1/\eps) + 5$. Since then, the study of non-malleable extractors has seen significant progress starting from the first explicit construction in \cite{DLWZ11}, with further connections to independent source extractors established in \cite{Li12b, Li13a, CZ15}. Previous to this work, the best known seeded non-malleable extractor is due to the author \cite{Li17}, which works for entropy $k \geq O(\log n+\log(1/\e)\log \log (1/\e))$ and has seed length $d= O(\log n+\log(1/\e)\log \log (1/\e))$.\ Although close to optimal, the extra $O(\log \log (1/\e))$ factor in the entropy requirement implies that by using this extractor, one can only get two-round privacy amplification protocols with optimal entropy loss for security parameter up to $s=\Omega(k/\log k)$. This still falls short of achieving the holy grail, and may be problematic for some applications. For example, even if the shared weak source has slightly super-logarithmic entropy, the error of the protocol can still be sub-polynomially large; while ideally one can hope to get negligible error, which is important for other cryptographic applications based on this.\ The only previous protocol that can achieve security parameter up to $s=\Omega(k)$ is the work of \cite{ckor}, which has entropy loss $O(\log n+s)$ but also uses $O(\log n+s)$ rounds of interactions, much larger than $2$. This also results in a total communication complexity of $O((\log n+s)^2)$ and requires the two parties' local random bits to be at least this long. 

\paragraph{Seedless non-malleable extractors and non-malleable codes.}
Seedless non-malleable extractors were first introduced by Cheraghchi and Guruswami \cite{CG14b} to study non-malleable codes \cite{DPW10}, a generalization of standard error correcting codes to handle a much larger class of attacks. Informally, a non-malleable code is defined w.r.t. a specific family of tampering functions $\cal F$. The code consists of a randomized encoding function $E$ and a deterministic decoding function $D$, such that for any $f \in \cal F$, if a codeword $E(x)$ is modified into $f(E(x))$, then the decoded message $x'=D(f(E(x)))$ is either the original message $x$ or a completely unrelated message.\ The formal definition is given in Section~\ref{sec:nmtext}. In \cite{DPW10}, Dziembowski et.\ al showed that such codes can be used generally in tamper-resilient cryptography to protect the memory of a device. 

Even with such generalization, non-malleable codes still cannot exist if $\cal F$ is completely unrestricted. However, they do exist for many broad families of tampering functions. One of the most studied families of tampering functions is the so called \emph{t-split-state} model. Here, a $k$-bit message $x$ is encoded into a codeword with $t$ parts $y_1, \cdots, y_t$, each of length $n$.\ An adversary can then arbitrarily tamper with each $y_i$ independently. In this case, the rate of the code is defined as $k/(tn)$. 

This model arises naturally in many applications, typically when different parts of memory are used to store different parts of $y_1, \cdots, y_t$. Such a code can also be viewed as a kind of ``non-malleable secret sharing scheme". The case of $t=2$ is the most useful and interesting setting, since $t=1$ corresponds to the case where $\cal F$ is unrestricted. Again, there has been a lot of previous work on non-malleable codes in this model. In this paper we will focus on the information theoretic setting.

Dziembowski et.\ al  \cite{DPW10} first proved the existence of non-malleable codes in the split-state model.\ Cheraghchi and Guruswami \cite{CG14a} showed that the optimal rate of such codes in the $2$-split-state model is $1/2$.\ Since then a major goal is to construct explicit non-malleable codes in the $2$-split-state model with constant rate.\ The first construction appears in \cite{DKO13}, with later improvements in \cite{ADL14, Agw14, ADKO15}, but all constructions only achieve rate $n^{-\Omega(1)}$. 

Cheraghchi and Guruswami \cite{CG14b} found a way to construct non-malleable codes in the $t$-split state model using non-malleable $t$-source extractors.\ Chattopadhyay and Zuckerman \cite{CZ14} constructed the first seedless non-malleable extractor, which works for $10$ independent sources with entropy $(1-\gamma)n$, and consequently they obtained a constant rate non-malleable code in the $10$-split-state model. Subsequently, constructions of non-malleable two source extractors appeared in \cite{CGL15} and \cite{Li17}. Both constructions work for min-entropy $k=(1-\gamma)n$, and the former gives a non-malleable code in the $2$-split state model with rate $n^{-\Omega(1)}$ while the latter achieves rate $\Omega(\frac{1}{\log n})$. Very recently, a work by Kanukurthi et.\ al \cite{KOS17} achieved constant rate in the $4$-split state model, and another one by Gupta et.\ al \cite{GMW18} achieved constant rate in the $3$-split state model, but the best construction in the $2$-split state model still only achieves rate $\Omega(\frac{1}{\log n})$ \cite{Li17}. 

As can be seen from the above discussions, extensive past research has established strong connections among these different topics, and provided solutions close to optimal. However, there still remains a gap and the motivation to close this gap remains strong.

\subsection{Our Results}
In this paper we introduce a set of new techniques to further push the frontier in the above questions. Our techniques lead to improvements in all the questions discussed, and in several cases partially optimal constructions.\ In contrast, all previous work only obtain close to optimal constructions.\ Our first theorem gives explicit seeded non-malleable extractors with optimal entropy requirement. 
\BT \label{thm1}
There exists a constant $C>1$ such that for any constant $a \in \N, a \geq 2$, any $n, k \in \N$ and any $0<\e<1$ with $k \geq C(\log \log n+a \log(1/\e))$, there is an explicit construction of a strong seeded $(k, \e)$ non-malleable extractor $\zo^n \times \zo^d \to \zo^m$ with $d=O(\log n)+\log(1/\e)2^{O(a(\log \log (1/\e))^\frac{1}{a})}$ and $m =\Omega(k)$. 
\ET

Note that this theorem provides a trade-off between the entropy requirement and the seed length. For example, if we take $a=2$, then the entropy requirement is $O(\log\log n+\log(1/\e))$ while the seed length is $O(\log n)+2^{O(\sqrt{\log \log (1/\e)})}\log(1/\e)=O(\log n)+\log^{1+o(1)}(1/\e)$. By a recent result of Gur and Shinkar \cite{GurS17}, the entropy requirement in our construction is asymptotically optimal. Combined with the protocol in \cite{DW09}, this gives the following theorem.

\BT
For any constant integer $a \geq 2$ there exists a constant $0<\alpha<1$ such that for any $n, k \in \N$ and security parameter $s \leq \alpha k$, there is an explicit two-round privacy amplification protocol with entropy loss $O(\log \log n+s)$,\ in the presence of an active adversary.\ The communication complexity of the protocol is $O(\log n)+s2^{O(a(\log s)^\frac{1}{a})}$.
\ET

Our two-round protocol has optimal entropy loss for security parameter up to $s=\Omega(k)$, thus achieving the holy grail of this problem.\ Compared to the $O(\log n+s)$-round protocol in \cite{ckor}, our protocol also has better dependence on $n$ and significantly better communication complexity.\ We remark that for clarity, the above theorem assumes an optimal seeded extractor with seed length $O(\log (n/\e))$ and entropy loss $O(\log(1/\e))$. However the best known extractor with such seed length \cite{GuruswamiUV09} has an additional entropy loss of $\alpha k$ ($\alpha$ is any constant), which will also appear in the protocol (the same as in \cite{DW09} and \cite{ckor}).\ Yet this is just an artifact of current seeded extractor constructions, and not of the protocol itself. Moreover, one can avoid this loss by using the extractor in \cite{rrv:all}, which has optimal entropy loss but larger seed length (e.g., $O(\log^2 n \log(1/\e) \log (k))$). This only affects the the communication complexity of the protocol.

We also remark that the $O(\log \log n)$ term in both theorems is also the best possible (up to constant) if one wants to apply the two-round protocol in \cite{DW09}. This is because the output of the non-malleable extractor is used in the second round as the key for a message authentication code (MAC) that authenticates the seed of a strong seeded extractor with security parameter $s$. Since the seed of the extractor uses at least $\Omega(\log n)$ bits, the MAC requires a key of length at least $\log \log n+s$. See \cite{DW09} for more details.
 
We can also achieve smaller seed length while requiring slightly larger entropy.

\BT \label{thm2}
There exists a constant $C>1$ such that for any $n, k \in \N$ and $0<\e<1$ with $k \geq C(\log \log n+\log(1/\e)\log \log \log(1/\e))$, there is an explicit construction of a strong seeded $(k, \e)$ non-malleable extractor $\zo^n \times \zo^d \to \zo^m$ with $d=O(\log n+\log(1/\e)(\log \log(1/\e))^2)$\footnote{The exponent $2$ can be reduced to be arbitrarily close to $\log 3$.} and $m =\Omega(k)$.
\ET

\BT
There exists a constant $0<\alpha<1$ such that for any $n, k \in \N$ and security parameter $s \leq \alpha k/ \log \log k$, there is an explicit two-round privacy amplification protocol with entropy loss $O(\log \log n+s)$, in the presence of an active adversary. The communication complexity of the protocol is $O(\log n+s \log^2 s)$.
\ET

\begin{remark}
In both Theorem~\ref{thm1} and Theorem~\ref{thm2}, the dependence on error $\e$ in the seed length and the entropy requirement can be switched. For example, in Theorem~\ref{thm1}, we can also achieve $k \geq C \log \log n+\log(1/\e)2^{C \cdot a(\log \log (1/\e))^\frac{1}{a}}$ and $d =O(\log n+a \log(1/\e))$.\ In other words, we can achieve asymptotically optimal parameters in either the seed length or the entropy requirement, but not in both. In addition, \end{remark}

\noindent We also have the following non-malleable two-source extractor and seeded non-malleable extractor.

\BT \label{thm3}
There exists a constant $0< \gamma< 1$ and a non-malleable two-source extractor for $(n, (1-\gamma)n)$ sources with error $2^{-\Omega(n \log \log n/\log n)}$ and output length $\Omega(n)$. 
\ET

\BT \label{thm4}
There is a constant $C>0$ such that for any $\e>0$ and $n, k \in \N$ with $k \geq C (\log \log n +\frac{\log(1/\e) \log \log (1/\e)}{\log \log \log(1/\e)})$, there is an explicit strong seeded non-malleable extractor for  $(n, k)$ sources with seed length $d = O (\log n +\frac{\log(1/\e) \log \log (1/\e)}{\log \log \log(1/\e)})$, error $\e$ and output length $\Omega(k)$. 
\ET

Combined with the techniques in \cite{BDT16}, we obtain the following theorem which gives improved constructions of two-source extractors and Ramsey graphs.

\BT \label{thm5}
For every constant $\e >0$, there exists a constant $C>1$ and an explicit two source extractor $\Ext: (\zo^n)^2 \to \zo$ for entropy $k \geq C\frac{\log n \log \log n}{\log \log \log n}$ with error $\e$. 
\ET

\BCR
For every large enough integer $N$ there exists a (strongly) explicit construction of a $K$-Ramsey graph on $N$ vertices with $K=(\log N)^{O(\frac{\log \log \log N}{\log \log \log \log N})}$. 
\ECR

For non-malleable codes in the $2$-split state model, we have the following theorem.

\BT 
There are constants $0< \eta, \mu<1$ such that for any $n \in \N$ and $2^{-\frac{\mu n}{\log n}} \leq \e \leq \eta$ there exists an explicit non-malleable code in the $2$-split-state model with block length $2n$, rate  $\Omega(\frac{\log \log \log (1/\e)}{\log \log (1/\e)})$ and error $\e$.
\ET

Note that if we choose $\e=2^{-c}$ for some constant $c>1$, then we get a non-malleable code with rate $\Omega(\frac{\log \log c}{\log c})$ and error $2^{-c}$.\ This gives the first construction of an explicit non-malleable code in the $2$-split-state model with \emph{constant} rate.\ Note that the error can be arbitrarily small, and the dependence of the rate on the error is pretty good. For example, even if one wants to achieve error $2^{-2^{100}}$, which is more than enough for any practical application, the rate is on the order of $1/16$. On the other hand, if we choose $\e=2^{-\polylog(n)}$, then we get a non-malleable code with negligible error and rate $\Omega(\frac{\log \log \log n}{\log \log n})$, which already improves the rate in \cite{Li17} exponentially. 

We can also achieve close to exponentially small error with an improved rate. 

\BT \label{thm6}
For any $n \in \N$ there exists a non-malleable code with efficient encoder/decoder in the $2$-split-state model with block length $2n$, rate  $\Omega(\log \log n/\log n)$ and error $\e=2^{-\Omega(n \log \log n/\log n)}$.
\ET

\subsection{Overview of The Constructions and Techniques}
We demonstrate our techniques here by an informal overview of our constructions.\ Throughout this section we will be mainly interested in the dependence of various parameters (e.g., seed length, entropy requirement) on the error $\e$, since this makes the presentation cleaner.\ The dependence on $n$ comes from the alternating extraction between the seed and the source, thus the seed needs to have an $O(\log n)$ term while the source only needs an $O(\log \log n)$ term. 

All recent constructions of non-malleable extractors essentially follow the same high level sketch: first obtain a small advice on $L=O(\log(1/\e))$ bits such that with probability $1-\e$, the advice is different from its tampered version. Then, use the rest of the inputs, together with a correlation breaker with advice (informally introduced in \cite{CGL15} and formally defined in \cite{Coh15nm}) to obtain the final output. There are several constructions of the correlation breaker, with the most efficient one using a non-malleable independence preserving merger ($\nipm$ for short, introduced in \cite{Coh16b} and generalized in \cite{CL16}). The $\nipm$ takes an $L \times m$ random matrix $V$ with $m=O(\log(1/\e))$ and use the other inputs to merge it into one output. It has the property that if the matrix has one row which is uniform given the corresponding row in its tampered version\footnote{Sometimes we also require the other rows to be uniform, in order to make the construction simpler. This is the case of this paper, but we ignore the issue here for simplicity and clarity.} (which can be obtained from the advice and inputs), then the output is guaranteed to be uniform given the tampered output. From now on, we assume the inputs to the extractor are two independent sources $X$ and $Y$ (in the case of a seeded non-malleable extractor, $Y$ can be viewed as the seed).

Previously, the best construction of an $\nipm$ is due to the author \cite{Li17}, which works roughly as follows. Suppose the matrix $V$ is a deterministic function of the source $X$, then we first generate $\ell =\log L$ random variables $(Y_1, \cdots, Y_{\ell})$ from $Y$, such that each $Y_i$ is close to uniform given the previous random variables and their tampered versions (i.e., $(Y_1, Y'_1, \cdots, Y_{i-1}, Y'_{i-1})$). We call this property the \emph{look-ahead} property. Next, we run a simple merger for $\ell$ iterations, with each iteration using a new $Y_i$ to merge every two consecutive rows in $V$, thus decreasing the number of rows by a factor of $2$. We output the final matrix $V$ which has one row. 

Let's turn to the entropy requirement.\ In this construction each $Y_i$ needs to have at least $\Omega(\log(1/\e))$ bits in order to ensure the error is at most $\e$, thus it is clear that $Y$ needs to have entropy at least $\Omega(\ell \log(1/\e))=\Omega(\log(1/\e) \log \log (1/\e))$. However, it turns out that $X$ also needs to have such entropy, for the following two reasons.\ First, in each iteration after we apply the simple merger, the length of each row in the matrix decreases by a constant factor (due to the entropy loss of any seeded extractor).\ Thus we cannot afford to just repeat the process for $\ell$ times since that would require the original row in $V$ (and hence $X$) to have entropy at least $\polylog(1/\e)$. Instead, we again create $\ell$ random variables $(X_1, \cdots, X_{\ell})$ from $X$ with the look-ahead property, and in each iteration after merging we use each row of the matrix to extract from a new $X_i$ (using a standard seeded extractor, and possibly after first extracting from another new $Y_i$), to restore the length of the rows in the matrix. We need the look-ahead property in $(X_1, \cdots, X_{\ell})$ and $(Y_1, \cdots, Y_{\ell})$ so that after each iteration we can fix the previously used random variables and maintain the independence of $X$ and $Y$, as well as the fact that the matrix is a deterministic function of $X$. Each $X_i$ again needs at least $\Omega(\log(1/\e))$ bits so this puts a lower bound on the entropy of $X$. 

Second, in order to prepare the random variables $(Y_1, \cdots, Y_{\ell})$, we in fact run an alternating extraction protocol between (part of) $X$ and $Y$. This protocol lasts $2 \ell$ rounds between $X$ and $Y$, and in each round either $X$ or $Y$ needs to spend $\Omega(\log(1/\e))$ random bits. This again puts a lower bound of $\Omega(\ell \log(1/\e))$ on the entropy of $X$.

We remark that the above description is slightly different from the standard definition of an $\nipm$, where the only input besides the matrix $V$ is $Y$. Indeed, in \cite{Li17} it was presented as a correlation breaker. However, these two objects are actually similar, and for this paper it is more convenient to consider $\nipm$s with an additional input $X$, which is independent of $Y$ but may be correlated with $V$. We will use this notion here and formally define it in Section~\ref{sec:nmipm}.

\paragraph{Improved merger construction.} We develop new techniques to break the above barriers. For the first problem, our key observation is that we can \emph{recycle} the entropy in $X$, similar in sprit to what has been done in previous constructions of pseudorandom generators for small space computation \cite{Nisan92, NisanZ96}. Indeed, the random variables $(X_1, \cdots, X_{\ell})$ can be replaced by the original source $X$, as long as we have slightly more (e.g., $2\ell$) $Y_i$'s and they satisfy the look ahead property. To achieve this we crucially use the property that the $\nipm$ only needs one row of $V$ to be uniform given the corresponding row in its tampered version, and does not care about the dependence among the rows of $V$ (they can have arbitrary dependence). Consider a particular iteration $i$ in which we have just finished applying the simple merger. We can first fix all random variables $\{Y_j\}$ that have been used so far, and conditioned on this fixing we know that $X$ and $Y$ are still independent, and the matrix $V$ is a deterministic function of $X$, which is independent of all random variables obtained from $Y$. To restore the length of each row in $V$, we use each row of $V$ to first extract $O(\log(1/\e))$ bits from $Y_{j+1}$, and then extract back from the original source $X$. Note that we only need to consider each row separately (since we don't care about the dependence among them). Assume row $h$ in $V$  has the property that $V_h$ is uniform given $V'_h$ (the tampered version). Since each random variable only has $O(\log(1/\e))$ bits, as long as the entropy of $X$ is $c\log(1/\e)$ for a large enough constant $c>1$, we can argue that conditioned on the fixing of $(V_h, V'_h)$, $X$ still has entropy at least some $O(\log(1/\e))$. On the other hand since  $V_h$ is uniform given $V'_h$, their corresponding outputs after extracting from $(Y_{j+1}, Y'_{j+1})$ will also preserve this independence; and conditioned on the fixing of $(V_h, V'_h)$, these outputs are deterministic functions of $(Y, Y')$, which are independent of $(X, X')$. Thus they can be used to extract back from $(X, X')$ and preserve the independence. By standard properties of a strong seeded extractor, this holds even conditioned on the fixing of $(Y_{j+1}, Y'_{j+1})$. Note that conditioned on the further fixing of $(Y_{j+1}, Y'_{j+1})$, the new matrix is again a deterministic function of $X$, thus we can go into the  next iteration. Therefore, by recycling the entropy in $X$, altogether we only need $X$ to have entropy some $O(\log(1/\e))$. In each iteration we use two new $Y_i$'s so we need roughly $2\ell$ such random variables.


However, we still need to address the second problem, where we need to generate the random variables $(Y_1, \cdots, Y_{2\ell})$.\ The old way to generate them by using an alternating extraction protocol requires entropy roughly $O(\ell \log(1/\e))$ from $X$.\ To solve this problem, we develop a new approach that requires much less entropy from $X$. For simplicity assume that $Y$ is uniform, we first take $2\ell$ slices $Y^i$ from $Y$, where $Y^i$ has size $(2^i-1)d$ for some $d=O(\log(1/\e))$. This ensures that even conditioned on the fixing of $(Y^1, Y'^1, \cdots, Y^{i-1}, Y'^{i-1})$, the (average) conditional min-entropy  of $Y_i$ is at least $(2^i-1)d-2 \cdot (2^{i-1}-1)d=d$. Then, we can take $O(\log(1/\e))$ uniform bits obtained from $X$, and use the \emph{same} bits to extract $Y_i$ from $Y^i$ for every $i$. As long as we use a strong seeded extractor here, we are guaranteed that $(Y_1, \cdots, Y_{2\ell})$ satisfy the look-ahead property; and moreover conditioned on the fixing of the $O(\log(1/\e))$ bits from $X$, we have that $(Y_1, \cdots, Y_{2\ell})$ is a deterministic function of $Y$. Note here again we only require entropy $O(\log(1/\e))$ from $X$, and together with the approach described above this gives us a non-malleable extractor where $X$ can have entropy $O(\log(1/\e))$. However $Y$ will need to have entropy at least $2^{2\ell} O(\log(1/\e))=O(\log^3(1/\e))$.

To improve the entropy requirement of $Y$, we note that in the above approach, we only used part of $X$ once to help obtaining the $\{Y^i\}$. Thus we have to use larger and larger slices of $Y$ which actually waste some entropy.\ Instead, we can use several parts of $X$, each with $O(\log(1/\e))$ uniform bits. For example, suppose that we have obtained $X^1$ and $X^2$, where each is uniform on some $O(\log(1/\e))$ bits and $X^2$ is uniform even conditioned on the fixing of $(X^1, X'^1)$.\ We can now take some $t$ slices $\{Y^i\}$ of $Y$, each of length $(2^i-1) \cdot 2d$ for some parameters $t, d$.\ We first use $X^1$ to extract from each $Y^i$ and obtain $d$ uniform bits. Note that conditioned on the fixing of $(X^1, X'^1)$, these $t$ random variables already satisfy the look-ahead property. Now for each of these $d$ bits obtained from $Y^i$, we can apply the same process, i.e., we take some $t$ slices of these $d$ bits, each of length $(2^i-1) \cdot O(\log(1/\e))$ and then use  $X^2$ to extract from each of them. This way we obtain $t^2$ random variables $\{Y_i\}$ that satisfy the look-ahead property.\ We can thus choose $t^2=2\ell$ which means $t=O(\sqrt{\ell})$. The entropy requirement of $Y$ is roughly $(2^t-1) \cdot (2^t-1) O(\log(1/\e))=O(2^{2t} \log(1/\e))=2^{O({\sqrt{\ell}})}\log(1/\e)$, while the entropy requirement for $X$ is $2 \cdot O(\log(1/\e))+O(\log(1/\e))=O(\log(1/\e))$. This significantly improves the entropy requirement of $Y$.

We can repeat the previous process and use some $a$ parts $(X^1, \cdots, X^a)$ obtained from $X$.\ As long as $a$ is a constant,\ $X$ only needs entropy $O(a \log(1/\e))=O( \log(1/\e))$, while the entropy requirement of $Y$ is reduced to $2^{O(a{\ell}^{\frac{1}{a}})}\log(1/\e)=2^{O(a{\log \log (1/\e)}^{\frac{1}{a}})}\log(1/\e)$.\ To prepare the $a$ parts of $X$, we perform an initial alternating extraction between $X$ and $Y$, which only needs entropy $O(a \log(1/\e))$ from either of them.\ This gives Theorem~\ref{thm1}.\ In the extreme case, we can try to minimize the entropy requirement of $Y$ by first creating $\log \ell+1=\log \log \log (1/\e)+O(1)$ $X^i$'s, and in each step using a new $X^i$ to double the number of $Y_i$'s. This can be done by using the same $X^i$ to do an alternating extraction of two rounds with each $Y_i$ in parallel. Thus after $\log \ell+1$ steps we obtain $(Y_1, \cdots, Y_{2\ell})$. Now $X$ needs to have entropy $O(\log (1/\e) \log \log \log (1/\e))$. Ideally, we would want to claim that $Y$ needs entropy $O(\log (1/\e) \log \log (1/\e))$, but due to technical reasons we can only show that this works as long as $Y$ has entropy $O(\log (1/\e) (\log \log (1/\e))^2)$.

\paragraph{The balanced case.} In the above discussion, the entropy requirement for $X$ and $Y$ is unbalanced, in the sense that one of them can be quite small, while the other is relatively large. For applications to two-source extractors and non-malleable codes, we need a balanced entropy requirement.\ Upon first look it does not seem that our new techniques can achieve any improvement in this case, since we are still merging two rows of the matrix $V$ in each step, and for this merging we need at least $\Omega(\log(1/\e))$ fresh random bits. Note that we need $\ell=\log L=\log \log(1/\e)$ steps to finish the merging, thus it seems the total entropy requirement is at least $\Omega(\log (1/\e) \log \log (1/\e))$.

Our key observation here is that we can again apply the idea of recycling entropy. Specifically, let us choose a parameter $t \in \N$ and we merge every $t$ rows in the matrix $V$ at each step, using some merger that we have developed above. For example, we can choose the merger which for merging $t$ rows, requires $X$ to have entropy $O(\log(1/\e))$ and $Y$ to have entropy $2^{O(\sqrt{\log t})}\log(1/\e)$. This will take us $\frac{\log L}{\log t}$ steps to finish merging, and we will do it in the following way. First, we create $s=O(\frac{\log L}{\log t})$ random variables $X_1, \cdots, X_{s}$ that satisfy the look-ahead property. Then, in each step of the merging, we will use a new $X_j$. The $X_j$'s can be prepared by taking a small slice of both $X$ and $Y$ and do an alternating extraction protocol with $O(s)$ rounds, which consumes entropy $O(s \log(1/\e))=O(\frac{\log L}{\log t} \log(1/\e))$ from both $X$ and $Y$. However, in each step of the merging, we will \emph{not} use fresh entropy from $Y$, but will recycle the entropy in $Y$. Note that by doing this, we are recycling the entropy in both $X$ and $Y$. The recycling in $X$ is done within each step of applying the small merger, while the recycling in $Y$ is done between these steps. 

Now, consider a particular step $i$ in the merging. Since we are using a new $X_j$ in each step, we can fix all previous $X_j$'s that have been used and their tampered versions. Conditioned on this fixing, the matrix $V$ obtained so far (and the tampered version $V'$) is a deterministic function of $Y$, therefore independent of $X$. We now want to claim that conditioned on the random variable $(V, V')$, $Y$ still has high entropy. If this is true then we can take a new $X_{j+1}$ and apply a strong seeded extractor to $Y$ using $X_{j+1}$ as the seed, and the extracted random bits (which are deterministic functions of $Y$ conditioned on the fixing of $X_{j+1}$) can be used for merging in the next step. Also note that to apply the merger, we can take yet another new $X_{j+2}$ and use each row of $V$ to extract from $X_{j+2}$ and create a matrix $W$. Conditioned on the fixing of $(V, V')$, we have that $(W, W')$ is a deterministic function of $(X, X')$ and therefore independent of $(Y, Y')$. Moreover the independence between corresponding rows in $(V, V')$ is preserved in $(W, W')$ (i.e., there is also a row in $W$ that is uniform given the corresponding row in $W'$). Thus now we can indeed apply the merger again to $W$ and the extracted random bits from $Y$, possibly together with a new $X_{j+3}$. Again, this is similar in spirit to what has been done in previous constructions of pseudorandom generators for small space computation \cite{Nisan92, NisanZ96}. 

The above idea indeed works, except for the following subtle point:\ in the first several steps of merging, the matrix $V$ can have many rows and the size of $V$ can be larger than the entropy of $Y$, unless $Y$ has entropy $\Omega(\log^2(1/\e))$.\ Thus conditioning on $(V, V')$ may cause $Y$ to lose all entropy.\ To get around this, we again use the fact that we only need one row in $V$ to be independent of the corresponding row in $V'$  (call this the good row), and does not care about the dependence between different rows.\ Thus in each step, we only need to condition on the fixing of the $t$ rows that we are merging (and their tampered versions).\ This ensures that if originally there is a good row in these $t$ rows, then after merging the output is also a good row in the new matrix.\ Thus, we only need the entropy of $Y$ to be $O(t \log(1/\e)) +2^{O(\sqrt{\log t})}\log(1/\e)+O(\frac{\log L}{\log t} \log(1/\e))=O(t \log(1/\e)+\frac{\log L}{\log t} \log(1/\e))$ since we will maintain the length of each row in $V$ to be $O(\log(1/\e))$. Now by choosing $t= \frac{\log L}{\log \log L}$, both $X$ and $Y$ only need entropy $O(\frac{\log L}{\log \log L} \log(1/\e))=O(\frac{ \log(1/\e)  \log \log (1/\e)}{ \log \log \log (1/\e)})$.\ By the connections in \cite{Li17, BDT16, CG14b}, this dependence gives Theorem~\ref{thm3}, \ref{thm4}, \ref{thm5} and \ref{thm6}. 

\paragraph{Non-malleable codes.} To further improve the rate of non-malleable codes in the $2$-split state model, we re-examine the connection between non-malleable codes and non-malleable two-source extractors found by Cheraghchi and Guruswami \cite{CG14b}. They showed that given a non-malleable two-source extractor with error $\e$ and output length $m$, the uniform sampling of the pre-image of any given output gives an encoding of a non-malleable code in the  $2$-split state model with error roughly $2^m \e$. This blow up of error comes from the conditioning on the event that the output of the extractor is a given string in $\bits^m$, which roughly has probability $2^{-m}$.\ Therefore, one needs $m < \log(1/\e)$, and thus the error of the extractor puts a limit on the rate of the code.

To break this barrier, we note that all recent constructions of non-malleable two-source extractors \cite{CGL15, Li17} follow a very special framework. As mentioned before, these constructions first obtain an advice $\tilde{\alpha}$ such that with probability $1-\e_1$ we have $\tilde{\alpha} \neq \tilde{\alpha}'$, where $\tilde{\alpha}'$ is the tampered version. Then, using a correlation breaker with advice one obtains the output. This part has error $\e_2$, and the final error of the extractor is $\e_1+\e_2$. 

In all previous work, this error is treated as a whole, but our key observation here is that these two errors $\e_1$ and $\e_2$ can actually be treated \emph{separately}. More specifically, the error that matters most for the rate of the code is actually $\e_2$, not $\e_1$.\ Intuitively, this is because the event $\tilde{\alpha} \neq \tilde{\alpha}'$ is determined by a set of random variables that have small size compared to the length of $X$ and $Y$. Thus even conditioned on the fixing of these random variables, $X$ and $Y$ still have plenty of entropy, which implies that the output of the extractor is still $\e_2$-close to uniform. Thus, as long as $\e_2$ is small, the output of the extractor is roughly independent of the event $\tilde{\alpha} \neq \tilde{\alpha}'$. Therefore, conditioned on any given output of the extractor, the event $\tilde{\alpha} \neq \tilde{\alpha}'$ still happens with probability roughly $1-\e_1$ and we won't be paying a price of $2^m \e_1$ here. Once this event happens, the correlation breaker ensures that the extractor is non-malleable with error $\e_2$, and we can use a similar argument as in \cite{CG14b} to get a non-malleable code with error roughly $2^m \e_2$. Thus the total error of the non-malleable code is roughly $\e_1+2^m \e_2$. Now, we just need $m < \log(1/\e_2)$.

We can now play with the two parameters $\e_1, \e_2$. The advice length $L$ is $\Omega(\log(1/\e_1))$ and we need to supply entropy $O(\frac{\log L}{\log \log L} \log(1/\e_2))$ by using our improved correlation breaker. If we can achieve $L=\Theta(\log(1/\e_1))$ then one can see that if we choose $\e_1$ to be any constant, then we can set $\e_2=2^{-\Omega(n)}$ and also $m=\Omega(n)$, thus we get a constant rate non-malleable code. If we set $\e_1=2^{-\polylog(n)}$ then we can set $\e_2=2^{-\Omega(\frac{n \log \log \log n}{\log \log n})}$ and thus we get rate $\Omega(\frac{\log \log \log n}{\log \log n})$. 

A technical issue here is how to achieve $L=\Theta(\log(1/\e_1))$ for any $\e_1$.\ In \cite{CGL15, Li17}, the advice is obtained by using some random seed $R$ to sample from an asymptotically good encoding of $X, Y$, and concatenating the sampled symbols with $R$. This puts a lower bound of $\log n$ on $L$, since we need at least this number of bits to sample from a string of length $n$. However this is not good enough to achieve constant rate. Our idea around this is to use repeated sampling. To illustrate the idea, suppose for example that we have obtained an advice $V$ such that $V \neq V'$ with probability $1-1/\poly(n)$ and $V$ has length $O(\log n)$. We now use another piece of independent random bits $R_1$ of length $O(\log \log n)$ to sample $O(\log \log n)$ bits from an asymptotically good encoding of $V$, and obtain a new advice $V_1$ by concatenating $R_1$ with the sample bits. This ensures that $V_1 \neq V_1'$ happens with probability $1-1/\polylog(n)$ conditioned on $V \neq V'$, and the length of $V_1$ is now $O(\log \log n)$. We repeat this process until we get the desired error  $\e_1$ (e.g., a constant) and the advice length is now $L=\Theta(\log(1/\e_1))$. Note that the total error is still $O(\e_1)$, the total number of random bits needed is small, and the process terminates in roughly $\log^*n$ steps. To prepare the independent random bits used in repeated sampling, we first take a small slice of $X$ and $Y$ and do an alternating extraction with roughly $\log^*n$ steps, which guarantees the bits used for sampling in later steps are independent of the previous ones and their tampered versions. Finally, some extra work are needed here to take care of the issue of fixed points, which is more subtle than \cite{CG14b} since now we are treating the two errors $\e_1$ and $\e_2$ separately.\\

\noindent{\bf Organization.}
The rest of the paper is organized as follows.\ We give some preliminaries in Section~\ref{sec:prelim}, and define alternating extraction in Section~\ref{sec:alt}.\ We present independence preserving mergers in Section~\ref{sec:nmipm}, correlation breakers in Section~\ref{sec:advcb}, non-malleable extractors in Section~\ref{sec:snmext}, and non-malleable codes in Section~\ref{sec:nmtext}. Finally we conclude with some open problems in Section~\ref{sec:open}.


\input{prelim.tex}

\input{reduction2.tex}

\input{ext.tex}

\input{newcode.tex}

\section{Discussion and Open Problems} \label{sec:open}
Several natural open problems remain here. The most intriguing one is how far we can push our new techniques. As mentioned above, one bottleneck here is that the computation of the merger is not a small space computation. If one can find a more succinct way to represent the computation, then it will certainly lead to further improvements (e.g., decrease the entropy requirement in two-source extractors to $O(\log n \sqrt{\log \log n})$). If in addition we can find a way to apply the recursive construction as in Nisan's generator \cite{Nisan92}, then it is potentially possible to decrease the entropy requirement in two-source extractors to $O(\log n \log \log \log n)$. We also believe our approach has the potential to eventually achieve truly optimal (up to constants) constructions. In addition, our techniques of treating the errors separately in non-malleable two-source extractors, may be useful in helping improve the rate of non-malleable codes for other classes of tampering functions (e.g., the affine tampering function and small depth circuits studied in \cite{CL17}).\\

\bibliographystyle{alpha}

\bibliography{refs}

\end{document}

%% file: prelim.tex
\section{Preliminaries} \label{sec:prelim}
We often use capital letters for random variables and corresponding small letters for their instantiations. Let $|S|$ denote the cardinality of the set~$S$. For $\ell$ a positive integer,
$U_\ell$ denotes the uniform distribution on $\zo^\ell$. When used as a component in a vector, each $U_\ell$ is assumed independent of the other components. When we have adversarial tampering, we use letters with prime to denote the tampered version of random variables. 
All logarithms are to the base 2.

\subsection{Probability Distributions}
\begin{definition} [statistical distance]Let $W$ and $Z$ be two distributions on
a set $S$. Their \emph{statistical distance} (variation distance) is
\begin{align*}
\Delta(W,Z) \eqdef \max_{T \subseteq S}(|W(T) - Z(T)|) = \frac{1}{2}
\sum_{s \in S}|W(s)-Z(s)|.
\end{align*}
\end{definition}

We say $W$ is $\eps$-close to $Z$, denoted $W \approx_\eps Z$, if $\Delta(W,Z) \leq \eps$.
For a distribution $D$ on a set $S$ and a function $h:S \to T$, let $h(D)$ denote the distribution on $T$ induced by choosing $x$ according to $D$ and outputting $h(x)$.

\BL \label{lem:sdis}
For any function $\alpha$ and two random variables $A, B$, we have $\Delta(\alpha(A), \alpha(B)) \leq \Delta(A, B)$.
\EL

\subsection{Average Conditional Min Entropy}
\label{avgcase}


\begin{definition}
The \emph{average conditional min-entropy} is defined as
\begin{align*}
 \thinf(X|W) &= - \log \left (\expect_{w \leftarrow W} \left [ \max_x \Pr[X=x|W=w] \right ] \right )
\\ &= - \log \left (\expect_{w \leftarrow W} \left [2^{-\hinf(X|W=w)} \right ] \right ).
\end{align*}
\end{definition}

\begin{lemma} [\cite{dors}]
\label{entropies}
For any $s > 0$,
$\Pr_{w \leftarrow W} [\hinf(X|W=w) \geq \thinf(X|W) - s] \geq 1-2^{-s}$.
\end{lemma}

\BL [\cite{dors}] \label{lem:amentropy}
If a random variable $B$ has at most $2^{\ell}$ possible values, then $\thinf(A|B) \geq \hinf(A)-\ell$.
\EL

\subsection{Prerequisites from Previous Work}

Sometimes it is convenient to talk about average case seeded extractors, where the source $X$ has average conditional min-entropy $\thinf(X|Z) \geq k$ and the output of the extractor should be uniform given $Z$ as well. The following lemma is proved in \cite{dors}.

\BL \cite{dors} \label{lem:avext}
For any $\delta>0$, if $\Ext$ is a $(k, \e)$ extractor then it is also a $(k+\log(1/\delta), \e+\delta)$ average case extractor.
\EL

For a strong seeded extractor with optimal parameters, we use the following extractor constructed in \cite{GuruswamiUV09}.

\BT [\cite{GuruswamiUV09}] \label{thm:optext} 
For every constant $\alpha>0$, there exists a constant $\beta>0$ such that for all positive integers $n,k$ and any $\e>2^{-\beta k}$, there is an explicit construction of a strong $(k,\e)$-extractor $\Ext: \bits^n \times \bits^d \to \bits^m$ with $d=O(\log n +\log (1/\e))$ and $m \geq (1-\alpha) k$. The same statement also holds for a strong average case extractor. 
\ET

\BT [\cite{ChorG88}] \label{thm:ip}
For every $0<m< n$ there is an explicit two-source extractor $\bip: \bits^n \times \bits^n \to \bits^m$ based on the inner product function, such that if $X, Y$ are two independent $(n, k_1)$ and $(n, k_2)$ sources respectively, then

\[(\bip(X, Y), X) \approx_{\e} (U_m, X) \text{ and } (\bip(X, Y), Y) \approx_{\e} (U_m, Y),\]
where $\e=2^{-\frac{k_1+k_2-n-m-1}{2}}.$
\ET

The following standard lemma about conditional min-entropy is implicit in \cite{NisanZ96} and explicit in \cite{MW97}.

\begin{lemma}[\cite{MW97}] \label{lem:condition} 
Let $X$ and $Y$ be random variables and let ${\calY}$ denote the range of $Y$. Then for all $\e>0$, one has
\[\Pr_Y \left [ H_{\infty}(X|Y=y) \geq H_{\infty}(X)-\log|{\calY}|-\log \left( \frac{1}{\e} \right )\right ] \geq 1-\e.\]
\end{lemma}

We also need the following lemma.

\BL \label{lem:jerror}\cite{Li13b}
Let $(X, Y)$ be a joint distribution such that $X$ has range $\calX$ and $Y$ has range $\calY$. Assume that there is another random variable $X'$ with the same range as $X$ such that $|X-X'| = \e$. Then there exists a joint distribution $(X', Y)$ such that $|(X, Y)-(X', Y)| = \e$.
\EL

%% file: reduction2.tex
\section{Alternating Extraction}\label{sec:alt}
Our constructions use the following alternating extraction protocol as a key ingredient. Alternating extraction was first introduced in \cite{DP07}, and has now become an important tool  in constructions related to extractors.

\begin{figure}[htb]
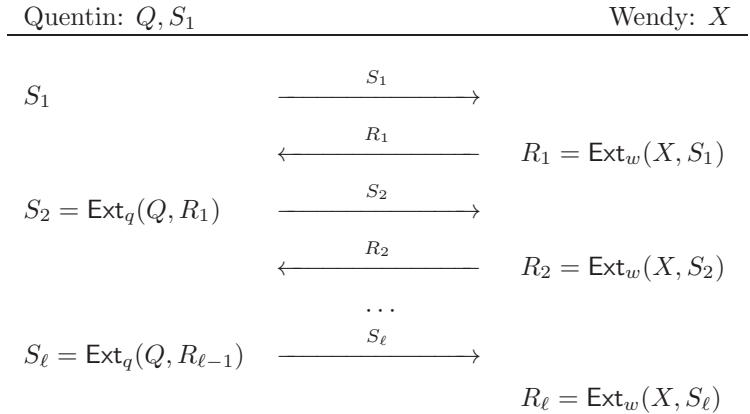

\begin{center}
\begin{small}
\begin{tabular}{l c l}
Quentin:  $Q, S_1$ & &~~~~~~~~~~Wendy: $X$ \\

\hline\\
$S_1$ & $\llrightarrow[\rule{2.5cm}{0cm}]{S_1}{} $ & \\
& $\llleftarrow[\rule{2.5cm}{0cm}]{R_1}{} $ & $R_1=\Ext_w(X, S_1)$ \\
$S_2=\Ext_q(Q, R_1)$ & $\llrightarrow[\rule{2.5cm}{0cm}]{S_2}{} $ & \\
& $\llleftarrow[\rule{2.5cm}{0cm}]{R_2}{} $ & $R_2=\Ext_w(X, S_2)$ \\
& $\cdots$ & \\
$S_\ell=\Ext_q(Q, R_{\ell-1})$ & $\llrightarrow[\rule{2.5cm}{0cm}]{S_\ell}{} $ & \\
& & $R_\ell=\Ext_w(X, S_\ell)$ \\
\hline
\end{tabular}
\end{small}
\caption{\label{fig:altext}
Alternating Extraction.
}
\end{center}
\end{figure}
\BD(Alternating Extraction)
 Assume that we have two parties, Quentin and Wendy. Quentin has a source $Q$,  Wendy has a source $W$. Also assume that Quentin has a uniform random seed $S_1$ (which may be correlated with $Q$). Suppose that $(Q, S_1)$ is kept secret from Wendy and $W$ is kept secret from Quentin.  Let $\Ext_q$, $\Ext_w$ be strong seeded extractors with optimal parameters, such as that in Theorem~\ref{thm:optext}. Let $r, s$ be two integer parameters for the protocol. For some integer parameter $\ell>0$, the \emph{alternating extraction protocol} is an interactive process between Quentin and Wendy that runs in $\ell$ steps. 

In the first step, Quentin sends $S_1$ to Wendy, Wendy computes $R_1=\Ext_w(W, S_1)$. She sends $R_1$ to Quentin and Quentin computes $S_2=\Ext_q(Q, R_1)$. In this step $R_1, S_2$ each outputs $r$ and $s$ bits respectively. In each subsequent step $i$, Quentin sends $S_i$ to Wendy, Wendy computes $R_i=\Ext_w(W, S_i)$. She replies $R_i$ to Quentin and Quentin computes $S_{i+1}=\Ext_q(Q, R_i)$. In step $i$, $R_i, S_{i+1}$ each outputs $r$ and $s$ bits respectively. Therefore, this process produces the following sequence: 

\begin{align*}
&S_1, R_1=\Ext_w(W, S_1), S_2=\Ext_q(Q, R_1), \cdots, \\ &S_\ell=\Ext_q(Q, R_{\ell-1}), R_\ell=\Ext_w(W, S_\ell).
\end{align*}
\ED

The output of an alternating extraction protocol is often described as a \emph{look-ahead extractor}, defined as follows. Let $Y=(Q, S_1)$ be a seed, the look-ahead extractor is defined as 

\[\laext(W,  Y)=\laext(W, (Q, S_1)) \eqdef R_1, \cdots, R_{\ell}.\]

The following lemma is a special case of Lemma 6.5 in \cite{CGL15}. 

\begin{lemma}\label{altext} Let $W$ be an  $(n_{w},k_{w})$-source and $W'$ be a random variable on $\{ 0,1\}^{n_w}$ that is arbitrarily correlated with $W$. Let $Y=(Q, S_1)$ such that $Q$ is a $(n_q,k_q)$-source,  $S_1$ is a uniform string on $s$ bits, and $Y' = (Q', S'_1)$ be a random variable arbitrarily correlated with $Y$, where $Q'$ and $S'_1$ are random variables on $n_q$ bits and $s$ bits respectively. Let $\Ext_q,\Ext_w$ be strong seeded extractors that extract $s$ and $r$ bits from sources with min-entropy $k$ with error $\epsilon$ and seed length $d \leq min\{r, s\}$. Suppose $(Y, Y')$ is independent of $(W,W')$, $k_q \ge k+ 2 (\ell-1) s+2 \log(\frac{1}{\epsilon})$, and $k_w \ge k+ 2 (\ell-1) r+2 \log(\frac{1}{\epsilon})$. Let $\laext$ be the look-ahead extractor defined above using $\Ext_q,\Ext_w$, and $(R_1, \cdots, R_{\ell})=\laext(W, Y)$, $(R'_{1}, \cdots, R'_{\ell})=\laext(W', Y')$. Then for any $0 \leq j \leq \ell-1$, we have

\begin{align*}
&(Y, Y',  \{R_{1}, R'_1, \cdots, R_{j}, R'_j\}, R_{j+1})  \\ \approx_{\e_1} &(Y, Y',  \{R_{1}, R'_1, \cdots, R_{j}, R'_j\}, U_r),
\end{align*}
where $\e_1=O(\ell \e)$.
\end{lemma}

\section{Non-Malleable Independence Preserving Merger}\label{sec:nmipm}
We now describe the notion of \emph{non-malleable independence preserving merger}, introduced in \cite{CL16} based on the notion of independence preserving merger introduced in \cite{Coh16b}. 

\BD A $(L, d', \eps)$-$\nipm: \zo^{Lm} \times \zo^d \rightarrow \zo^{m_1}$ satisfies the following property.  Suppose
\begin{itemize}
\item $\X,\X'$ are random variables, each supported on boolean $L\times m$ matrices s.t for any $i \in [L]$, $\X_i = U_m$,
\item $\{\Y,\Y'\}$ is independent of $\{ \X,\X'\}$, s.t $\Y,\Y'$ are each supported on $\zo^{d}$ and $H_{\infty}(\Y) \ge d'$,
\item there exists an $h \in [L]$ such that $(\X_h,\X'_h)=(U_m,\X'_h)$,
\end{itemize}
then 
\begin{align*}
|&(L,d', \eps)\text{-}\nipm(\X,\Y), (L,d', \eps)\text{-}\nipm(\X', \Y') \\ & -U_{m_1},  (L,d', \eps)\text{-}\nipm(\X', \Y')| \le \epsilon.
\end{align*}
\ED

We have the following construction and theorem.

\textbf{$L$-Alternating Extraction} We extend the previous alternating extraction protocol by letting Quentin have access to $L$ sources $Q_1,\ldots,Q_L$ (instead of just $Q$) which have the same length.  Now in the $i$'th round of the protocol, he uses $Q_i$ to produce the r.v $S_i=\Ext_{q}(Q_i,R_i)$. More formally, the following sequence of r.v's is  generated: $ S_1, R_1 = \Ext_{w}(W,S_1), S_2 = \Ext_{q}(Q_2, R_1),\ldots,R_{L-1}=\Ext_{w}(W, S_{\ell-1}),S_{L} = \Ext_{q}(Q_{L},R_{L-1})$. 

The $\nipm$ is now constructed as follows. Let $S_1$ be a slice of $\X_1$ with length $O(\log(d/\eps))$, then run the $L$-alternating extraction described above with $(Q_1,\ldots,Q_L)=(\X_1, \ldots, \X_L)$ and $W=\Y$. Finally output $S_L$.

\begin{theorem}[\cite{CL16}] \label{thm:nipm} 
There exists a constant $c>0$ such that for all integers $m,d, d', L>0$  and any $\epsilon>0$,  with $m \ge 4c L \log(d/\epsilon)$, $d' \ge 4c L \log(m/\epsilon)$, the above construction $\nipm:(\zo^{m})^{\ell} \times \zo^d \rightarrow \zo^{m_1}$ has output length $m_1 \geq 0.2 m$, such that if the following conditions hold:

\begin{itemize}
\item $\X,\X'$ are random variables, each supported on boolean $L\times m$ matrices s.t for any $i \in [L]$, $\X_i = U_m$,
\item $\{\Y,\Y'\}$ is independent of $\{ \X,\X'\}$, s.t $\Y,\Y'$ are each supported on $\zo^{d}$ and $H_{\infty}(\Y) \ge d'$,
\item there exists an $h \in [L]$ such that $(\X_h,\X'_h)=(U_m,\X'_h)$,
\end{itemize}
then 
\begin{align*}
| \nipm(\X, \Y), \nipm(\X', \Y'), \Y, \Y' -U_{m_1}, \nipm(\X', \Y'), \Y,\Y'|  \le L \epsilon.
\end{align*}
\end{theorem}

It is sometimes more convenient to consider $\nipm$s which use an additional source $X$ in the computation. We generalize the above definition as follows.

\BD A $(L, d, d', \eps)$-$\nipm: \zo^{Lm} \times \zo^d \times \zo^{d'} \rightarrow \zo^{m_1}$ satisfies the following property.  Suppose
\begin{itemize}
\item $V,V'$ are random variables, each supported on boolean $L\times m$ matrices s.t for any $i \in [L]$, $V_i = U_m$,
\item there exists an $h \in [L]$ such that $(V_h,V'_h)=(U_m,V'_h)$,
\item $\X, \X'$ are random variables, each supported on $d$ bits, such that $\X$ is uniform conditioned on $(V, V')$, 
\item $(\Y,\Y')$ is independent of $( V, V', \X,\X' )$, s.t $\Y,\Y'$ are each supported on $\zo^{d'}$ and $\Y$ is uniform,
\end{itemize}
If the function is an $\nipm$ that is strong in $\Y$ then
\begin{align*}
|&(L,d, d', \eps)\text{-}\nipm(V, \X, \Y), (L,d, d', \eps)\text{-}\nipm(V', \X', \Y'), \Y,\Y' \\ & -U_{m_1},  (L,d, d', \eps)\text{-}\nipm(V', \X', \Y'), \Y,\Y' | \le \epsilon.
\end{align*}

If the function is an $\nipm$ that is strong in $\X$ then

\begin{align*}
|&(L,d, d', \eps)\text{-}\nipm(V, \X, \Y), (L,d, d', \eps)\text{-}\nipm(V', \X', \Y'), \X,\X' \\ & -U_{m_1},  (L,d, d', \eps)\text{-}\nipm(V', \X', \Y'), \X,\X' | \le \epsilon.
\end{align*}
\ED

We will now use the above construction to give another $\nipm$, which recycles the entropy. Specifically, we have the following construction.

\begin{construction} \label{con:nnipm} Asymmetric $\nipm$.

Inputs:
\begin{itemize}
\item $L, m, n, d \in \N$ and an error parameter $\e>0$ such that $m \geq c\log (d/\e)$ and $d \geq c\log (n/\e)$ for some constant $c>1$.
\item A random variable $V$ supported on a boolean $L\times m$ matrix. 
\item An $(n, 6m)$ source $\X$.
\item Random variables $\Y_1,  \cdots, \Y_\ell$ where $\ell=\log L$ and each $\Y_i$ is supported on $\zo^d$. 
\end{itemize}

Output: a random variable $\W \in \zo^{m}$.\\ 

Let $V^0=V$. For $i=1$ to $\log L$ do the following.
\begin{enumerate}
\item Take a slice $\Y^1_i$ of $\Y_i$ with length $d/3$. Merge every two rows of $V^{i-1}$, using $\Y^1_i$ and the $\nipm$ from Theorem~\ref{thm:nipm}. That is, for every $j \leq t/2$ where $t$ is the current number of rows in $V^{i-1}$ (initially $t=L$), compute $\overline{V^{i-1}_j}=\nipm((V^{i-1}_{2j-1}, V^{i-1}_{2j}), \Y^1_i)$.

\item For every $j \leq t/2$, compute $\overline{\Y_{ij}}=\Ext_1(\Y_{i}, \overline{V^{i-1}_j})$, where $\Ext_1$ is the extractor in Theorem~\ref{thm:optext} and output $d/4$ bits.

\item For every $i \leq t/2$, compute $\widetilde{V^{i-1}_j}=\Ext_2(\X, \overline{\Y_{ij}})$, where $\Ext_2$ is the extractor in Theorem~\ref{thm:optext} and output $m$ bits.

\item Let $V^i$ with the concatenation of $\widetilde{V^{i-1}_j}, j =1, \cdots, t/2$. Note that the number of rows in $V^i$ has decreased by a factor of $2$.
\end{enumerate}

Finally output $\W=V^{\log L}$.
\end{construction}

\BL  \label{lem:nnipm} 
There is a constant $c>1$ such that suppose we have the following random variables:
\begin{itemize}
\item $V,V'$, each supported on a boolean $L\times m$ matrix s.t for any $i \in [L]$, $V_i = U_m$. In addition, there exists an $h \in [L]$ such that $(V_h,V'_h)=(U_m,V'_h)$.
 \item $\X, \X'$ where $\X$ is an $(n, 6m)$ source.
 \item Random variables $(\Y_1, \Y'_1), \cdots, (\Y_\ell, \Y'_{\ell})$ obtained from $\Y, \Y'$ deterministically, where $\ell=\log L$. These random variables satisfy the following look-ahead condition: $\forall j < \ell$, we have 

\[(\Y_j, \Y_1, \Y'_1, \cdots, \Y_{j-1}, \Y'_{j-1}) =(U_d, \Y_1, \Y'_1, \cdots, \Y_{j-1}, \Y'_{j-1}).\] In addition, $(V, V', \X, \X')$ is independent of $(\Y, \Y')$.
 \end{itemize}
 Let $\W$ be the output of the $\nipm$ on $(V, \X, \Y_1,  \cdots, \Y_\ell)$ and $\W'$ be the output of the $\nipm$ on $(V', \X', \Y'_1,  \cdots, \Y'_\ell)$. Then 
 
 \[(\W, \W', \Y, \Y') \approx_{O(L \e)} (U_{m}, \W', \Y, \Y').\]
\EL

\begin{proof}
We use induction to show the following claim.

\BCM
For every $0 \leq i \leq \ell=\log L$, the following holds after step $i$.
\begin{itemize}
\item $V^i,V'^i$ are each supported on boolean $(t=L/2^{i}) \times m$ matrices s.t for any $j \in [t]$, $(V^i_j, \Y, \Y') \approx_{\e_j} (U_m,\Y, \Y')$. In addition, there exists an $h \in [t]$ such that $(V^i_h,V'^i_h, \Y, \Y') \approx_{\e_i} (U_m,V'^i_h, \Y, \Y')$. Here $\e_i$ is the error after step $i$ which satisfies that $\e_0=0$ and $\e_{i+1} \leq 2 \e_i +4\e$.
\item Conditioned on the fixing of $\Y_1, \Y'_1, \cdots, \Y_{j}, \Y'_{j}$, each of $V^i$ and $V'^i$ is a deterministic function of $V, V', \X, \X'$.
\end{itemize}
\ECM

For the base case of $i=0$, the claim clearly holds. Now assume that the claim holds for $i$, we show that it holds for $i+1$. 

We first fix $\Y_1, \Y'_1, \cdots, \Y_{i}, \Y'_{i}$. By the induction hypothesis, conditioned on the fixing of these random variables, each of $V^i$ and $V'^i$ is a deterministic function of $V, V', \X, \X'$, and thus independent of $(\Y_{i+1}, \Y'_{i+1})$. We only consider the row $h \in [t]$ such that $(V_h,V'_h) \approx_{4 \cdot 2^i \e} (U_m, V'_h)$, since the analysis for the rest of the rows are similar and simpler. 

First we ignore the error $\e_i$. By Theorem~\ref{thm:nipm}, and note that we are merging every two rows at one step, we can choose a suitable constant $c>1$ in the construction such that 

\[(\overline{V^i_{h'}}, \overline{V'^i_{h'}}, \Y^1_{i+1}, \Y'^1_{i+1}) \approx_{2\e} (U_{m_1}, \overline{V'^i_{h'}}, \Y^1_{i+1}, \Y'^1_{i+1}),\]

where $h'=\lceil \frac{h}{2} \rceil$ and $m_1=0.2m$. We now fix $(\Y^1_{i+1}, \Y'^1_{i+1})$. Note that conditioned on the fixing, $\Y_{i+1}$ still has average conditional min-entropy at least $d-d/3=2d/3$ and is independent of $(\overline{V^i_{h'}}, \overline{V'^i_{h'}})$. Now we can first fix $\overline{V'^i_{h'}}$ and then $\overline{\Y'_{ih'}}$. Note that conditioned on this fixing, $\overline{V^i_{h'}}$ is still (close to) uniform and the average conditional min-entropy of $\Y_{i+1}$ is at least $2d/3-d/4>d/3$. Thus as long as $c$ is large enough, by Theorem~\ref{thm:optext} we have that

\[(\overline{\Y_{ih'}}, \overline{V^i_{h'}}) \approx_{\e} (U_{d/4}, \overline{V^i_{h'}}).\]

We now further fix $\overline{V^i_{h'}}$. Note that conditioned on this fixing, $\overline{\Y_{ih'}}$ is still (close to) uniform. Moreover conditioned on all the random variables we have fixed, $\overline{\Y}_{ih'}$ is a deterministic function of $\Y_1, \Y'_1, \cdots, \Y_{i+1}, \Y'_{i+1}$ and thus independent of $\X, \X'$. Also conditioned on all the random variables we have fixed, the average conditional min-entropy of $\X$ is at least $6m-2m_1 > 5m$. 

We can now further fix $\widetilde{V'^i_{h'}}$, which is a deterministic function of $\X'$. Conditioned on this fixing the independence of random variables still holds, while the average conditional min-entropy of $\X$ is at least $5m-m =4m$. Therefore by Theorem~\ref{thm:optext} we have that

\[(\widetilde{V^i_{h'}}, \overline{\Y_{ih'}}) \approx_{\e} (U_{m}, \overline{\Y_{ih'}}).\]

Since we have already fixed $\overline{\Y'_{ih'}}$ and $\widetilde{V'^i_{h'}}$, and note that conditioned on this fixing, $(\Y, \Y')$ are independent of $\widetilde{V^i_{h'}}$ which is a deterministic function of $\X$, we also have that

\[(\widetilde{V^i_{h'}}, \widetilde{\X'_{h'}}, \Y, \Y') \approx_{\e} (U_{m}, \widetilde{\X'_{h'}}, \Y, \Y').\]
Adding back all the errors we get that there exists an $h' \in [t]$ such that

\[(\widetilde{\X_{h'}},\widetilde{V'^i_{h'}}, \Y, \Y') \approx_{\e_{i+1}} (U_{m}, \widetilde{V'^i_{h'}}, \Y, \Y'),\]
where $\e_{i+1} \leq 2 \e_i +4\e$. Furthermore, it is clear that conditioned on the fixing of $\Y_1, \Y'_1, \cdots, \Y_{i+1}, \Y'_{i+1}$, each of $V^{i+1}$ and $V'^{i+1}$ is a deterministic function of $V, V', \X, \X'$.

We can now estimate the final error to be $\e_{\ell} \leq 4 (\sum_{i=1}^{\ell} 2^i \e)=O(L \e)$. Finally, when the number or rows in $V^i$ decreases to 1 after step $\ell$, the output $\W=V^{\log L}$ satisfies the conclusion of the lemma.
\end{proof}

We will now construct another $\nipm$. First we need the following lemma.

\BL \label{lem:convert}
For any constant $a \in \N$, any $\ell, s \in \N$ and any $\e>0$ there exists an explicit function $\conv_a: \bits^n \times \bits^{a \cdot d} \to \bits^{\ell \cdot s}$ with $d=O(\log(n/\e))$ and $n=2^{O(a \cdot \ell^{\frac{1}{a}})} \cdot s$ such that the following holds. Let $(Y, Y')$ be two random variables each on $n$ bits, and $Y$ is uniform. Let $(X=(X_1, \cdots, X_a), X'=(X'_1, \cdots, X'_a))$ be random variables each on $a \cdot d$ bits, where each $X_i$ and $X'_i$ is on $d$ bits. Further assume that $(X, X')$ satisfies the following look-ahead property: $\forall i \in [a]$, we have 

\[(X_{i}, X_1, X'_1, \cdots, X_{i-1}, X'_{i-1})=(U_d, X_1, X'_1, \cdots, X_{i-1}, X'_{i-1}).\]

Let $(W_1, \cdots, W_\ell)=\conv_a(Y, X)$ and $(W'_1, \cdots, W'_\ell)=\conv_a(Y', X')$. Then we have 

\[(X, X', W_1, W'_1, \cdots, W_\ell, W'_\ell) \approx_{O(\ell \e)} (X, X', U_s, W'_1, \cdots, U_s, W'_\ell),\]

where each $U_s$ is independent of previous random variables but may depend on later random variables.
\EL

\begin{proof}
We will prove the lemma by induction on $a$. For the base case $a=1$, consider the following construction. For $j=1, \cdots, \ell$, let $Y_j$ be a slice of $Y$ with length $(2^j-1) \cdot 2s$ (this is possible since the total entropy required is at most $2^{\ell} \cdot 2s$), and compute $W_j=\Ext(Y_j, X_1)$. Note that for any $j \in [\ell]$, conditioned on the fixing of $Y_1, Y'_1, \cdots, Y_{j-1}, Y'_{j-1}$, the average conditional min-entropy of $Y_j$ is at least $(2^j-1) \cdot 2s - 2 (2^{j-1}-1) \cdot 2s =2s$. Thus by Theorem~\ref{thm:optext} we have that

\[(W_j, Y_1, Y'_1, \cdots, Y_{j-1}, Y'_{j-1}, X, X') \approx_{\e} (U_s, Y_1, Y'_1, \cdots, Y_{j-1}, Y'_{j-1}, X, X').\]

Since $(W_1, W'_1, \cdots, W_{j-1}, W'_{j-1})$ is a deterministic function of $(Y_1, Y'_1, \cdots, Y_{j-1}, Y'_{j-1})$ and $(X, X')$, we also have that

\[(W_j, W_1, W'_1, \cdots, W_{j-1}, W'_{j-1}, X, X') \approx_{\e} (U_s, W_1, W'_1, \cdots, W_{j-1}, W'_{j-1}, W, W').\]

By adding all the errors the statement of the lemma holds.

Now assume that the lemma holds for $a$, we will construct another function $\conv_{a+1}$ for the case of $a+1$. First choose a parameter $t \in \N$ to be decided later. For $j=1, \cdots, \ell/t$, let $Y_j$ be a slice of $Y$ with length $(2^j-1) \cdot 2m$, where $m$ is the length of $Y$ (i.e., $n$) for $\conv_{a}$ when choosing $\ell=t$. Thus we have $m=2^{O(a \cdot t^{\frac{1}{a}})} \cdot s$. Now, for every $j$ we first use $X_1$ to compute $\hat{W}_j=\Ext(Y_j, X_1)$ and output $m$ bits, then compute $(\hat{W}_{1j}, \cdots, \hat{W}_{tj})=\conv_a(\hat{W}_j, X_2, \cdots, X_{a+1})$.  The final outputs are obtained by combining all the $\{\hat{W}_{ij}\}$ in sequence.

Note that by the same argument as above, we have that 

\[(X_1, X'_1, \hat{W}_1, \hat{W'}_1, \cdots, \hat{W}_{\ell/t}, \hat{W'}_{\ell/t}) \approx_{O(\frac{\ell}{t} \e)} (X_1, X'_1, U_m, \hat{W'}_1, \cdots, U_m, \hat{W'}_{\ell/t}).\]

Now we can fix $(X_1, X'_1)$. Note that conditioned on the fixing, $(\hat{W}_1, \hat{W'}_1, \cdots, \hat{W}_{\ell/t}, \hat{W'}_{\ell/t})$ is a deterministic function of $(Y, Y')$, thus independent of $(X, X')$. Now we can used the induction hypothesis to conclude that the statement holds for the case of $a+1$. Note that the total error is $O(\frac{\ell}{t} \e)+ \ell/t \cdot O(t \e)=O(\ell \e)$ since the part of $O(\frac{\ell}{t} \e)$ decreases as a geometric sequence. Finally, the entropy requirement of $Y$ is $(2^{\ell/t}-1) \cdot 2m = (2^{\ell/t}-1) \cdot 2 \cdot 2^{O(a \cdot t^{\frac{1}{a}})} \cdot s=2^{l/t+O(a \cdot t^{\frac{1}{a}})+1} \cdot s$.

We now just need to choose a $t$ to minimize this quantity. We can choose $t=\ell^{\frac{a}{a+1}}$ so that the entropy requirement of $Y$ is $2^{O((a+1) \cdot \ell^{\frac{1}{a+1}})} \cdot s$.
\end{proof}

We now have the following construction.

\begin{construction} \label{con:nnipm2} $\nipm_x$ (which is strong in $Y$) or $\nipm_y$ (which is strong in $X$).

Inputs:
\begin{itemize}
\item An error parameter $\e>0$ and a constant $a \in \N$.
\item A random variable $V$ supported on a boolean $L\times m$ matrix. 
\item A uniform string $X$ on $d_1$ bits.
\item A uniform string $Y$ on $d_2$ bits. 
\item Let $d= c\log (max\{d_1, d_2\}/\e)$ for some constant $c>1$. 

\end{itemize}

Output: $\nipm_x$ outputs a random variable $\W_x \in \zo^{m}$, and $\nipm_y$ outputs $\W_y \in \zo^{d}$. 

\begin{enumerate}

\item Let $\ell=\log L$.\footnote{Without loss of generality we assume that $L$ is a power of $2$. Otherwise add $0$ to the string until the length is a power of $2$.} Let $X_0$ be a slice of $X$ with length $4a \cdot d$, and $Y_0$ be a slice of $Y$ with length $4a \cdot d$. Use $X_0$ and $Y_0$ to run an alternating extraction protocol, and output $(R_0, \cdots, R_a)=\laext(X_0, Y_0)$ where each $R_i$ has $d$ bits. 

\item Compute $Z=\Ext(Y, R_0)$ and output $d_2/2$ bits, where $\Ext$ is the strong seeded extractor from Theorem~\ref{thm:optext}.

\item For every $i \in [L]$, compute $\overline{V_i}=\Ext(Y_0, V_i)$ and output $d$ bits. Then, compute  $\hat{V_i}=\Ext(X, \overline{V_i})$ and output $m$ bits.

\item Compute $(Z_1, \cdots, Z_{\ell})=\conv_a(Z, R_1, \cdots, R_a)$ where each $Z_i$ has $d$ bits. 

\item $\nipm_x$ outputs $\W_x=\nipm(\hat{V}, Z_1, \cdots, Z_{\ell})$, where $\nipm$ is the merger in Construction~\ref{con:nnipm} and Lemma~\ref{lem:nnipm}.  $\nipm_y$ outputs $\W_y =\Ext(Y, \W_x)$ with $d$ bits.
\end{enumerate}

\end{construction}

We now have the following lemma.

\BL \label{lem:nnipm2}
There exist a constant $c >1$ such that for any $\e>0$ and any $L, m, d_1, d_2, n \in \N$ such that $d \geq c (\log \max\{d_1, d_2\}+\log(1/\e))$, $m \geq d$, $d_1 \geq 8a \cdot d+6m$ and $d_2 \geq 8a \cdot d+c^{a \cdot \log ^{\frac{1}{a}} L} \cdot d$, the above construction gives an $(L, d_1, d_2, O(L \e))$-$\nipm$ that is either strong in $X$ or strong in $Y$. 
\EL

\begin{proof}
Note that $Y_0$ has min-entropy $4ad \geq 4d$, thus by Theorem~\ref{thm:optext} we have that for every $i \in [L]$, 

\[(\overline{V_i}, V_i) \approx_{\e} (U_d, V_i),\]

and there exists an $h \in [L]$ such that

\[(\overline{V_h}, \overline{V'_h}, V_h, V'_h) \approx_{\e} (U_d,  \overline{V'_h}, V_h, V'_h).\]

Note that conditioned on the fixing of $(V, V')$, we have that $(X, X')$ and $(Y, Y')$ are still independent, and furthermore $(\overline{V}, \overline{V'})$ is a deterministic function of $(Y, Y')$. Note that conditioned on the fixing of $(X_0, X'_0)$, the average conditional min-entropy of $X$ is at least $8a \cdot d+6m-2 \cdot 4a \cdot d=6m$. Thus again  by Theorem~\ref{thm:optext} we have that for every $i \in [L]$,

\[(\hat{V}_i, \overline{V_i}) \approx_{\e} (U_d, \overline{V_i}),\]

and there exists an $h \in [L]$ such that

\[(\hat{V}_h, \hat{V'}_h, \overline{V_h}, \overline{V'_h}) \approx_{\e} (U_d,  \hat{V'}_h, \overline{V_h}, \overline{V'_h}).\]

Note that now conditioned on the fixing of $(\overline{V_h}, \overline{V'_h})$, we have that $(X, X')$ and $(Y, Y')$ are still independent, and furthermore $(\hat{V}_h, \hat{V'}_h)$ is a deterministic function of $(X, X')$. Thus we basically have that conditioned on the fixing of $(X_0, X'_0, Y_0, Y'_0)$, $(\hat{V}, \hat{V'})$ is a deterministic function of $(X, X')$ and they satisfy the property needed by an $\nipm$.

Now, by Lemma~\ref{altext}, we have that 

\[( Y_0, Y'_0, R_0, R'_0, \cdots, R_a, R'_a) \approx_{O(a^2 \e)} ( Y_0, Y'_0, U_d, R'_0, \cdots, U_d, R'_a).\]

Note that conditioned on the fixing of $(Y_0, Y'_0)$, we have that $(X, X')$ and $(Y, Y')$ are still independent, and furthermore $(R_0, R'_0, \cdots, R_a, R'_a)$ is a deterministic function of $(X, X')$. Also the average conditional min-entropy of $Y$ is at least $d_2-2 \cdot 4a \cdot d =c^{a \cdot \log ^{\frac{1}{a}} L} \cdot d > 3d_2/4$ for a large enough constant $c$. Thus by Theorem~\ref{thm:optext} we have that 

\[(Z, R_0) \approx_{\e} (U_{d_2/2}, R_0).\]

We can now fix $(R_0, R_0)$. Note that now $(Z_0, Z'_0)$ is a deterministic function of $(Y, Y')$, and $d_2/2 > \frac{1}{2}c^{a \cdot \log ^{\frac{1}{a}} L} \cdot d$. Note that now $(R_1, R'_1, \cdots, R_a, R'_a)$ still satisfies the look-ahead property. Thus as long as $c$ is large enough, by Lemma~\ref{lem:convert} we have that

\[(Z_1, Z'_1, \cdots, Z_\ell, Z'_\ell, X_0, X'_0) \approx_{O(\ell \e)} (U_d, W'_1, \cdots, U_d, W'_\ell, X_0, X'_0).\]

We can now fix $(X_0, X'_0)$, and note that conditioned on this fixing $(Z_1, Z'_1, \cdots, Z_\ell, Z'_\ell)$ is a deterministic function of $(Y, Y')$. In summary, conditioned on the fixing of $(X_0, X'_0, Y_0, Y'_0)$, we have that $(\hat{V}, \hat{V'})$and $(Z_1, Z'_1, \cdots, Z_\ell, Z'_\ell)$ satisfy the conditions required by Lemma~\ref{lem:nnipm}. Therefore we can now apply that lemma to finish the proof. The total error is at most $O(L \e)+O(a^2 \e)+O(\e)+O(\ell \e)=O(L \e)$.
\end{proof}

The extreme case of the above construction gives the following $\nipm$.

\begin{construction} \label{con:nnipm3} $\nipm_x$ (which is strong in $Y$) or $\nipm_y$ (which is strong in $X$).

Inputs:
\begin{itemize}
\item An error parameter $\e>0$.
\item A random variable $V$ supported on a boolean $L\times m$ matrix. 
\item A uniform string $\X$ on $n$ bits.
\item A uniform string $\Y$ on $n'$ bits. 
\end{itemize}

Output: $\nipm_x$ outputs a random variable $\W_x \in \zo^{m}$, and $\nipm_y$ outputs $\W_y \in \zo^{O(\log(n/\e))}$. 

\begin{enumerate}
\item Let $d_1=c\log(n'/\e)$ and $d_2=c \log(n/\e)$. Take a slice $\X_0$ of $\X$ with length $10 \log \log L \cdot d_1$, and a slice $\Y_0$ of $\Y$ with length $10 \log \log L \cdot d_2$. 

\item Use $\X_0$ and $\Y_0$ to do an alternating extraction protocol, and output $(R_0, R_1, \cdots, R_{t})=\laext(\X_0, \Y_0)$ where $t=\log \log L$ and each $R_i$ has $4d_1$ bits, each $S_i$ (used in the alternating extraction) has $d_2$ bits. 

\item For each $i \in [L]$, compute $\overline{\Y}_i=\Ext(\Y_0, V_i)$ where each $\overline{\Y}_i$ outputs $d_2$ bits. Then compute $\overline{V}_i=\Ext(\X, \overline{\Y}_i)$ where each $\overline{V}_i$ outputs $m$ bits. Here $\Ext$ is the strong seeded extractor from Theorem~\ref{thm:optext}. Let $\overline{V}$ be the matrix whose $i$'th row is  $\overline{V}_i$. 


\item Let $\Y^0_1=\Y$. For $j=0$ to $\log \log L$ do the following. For $h =1$ to $2^j$, use $\Y^j_h$ and $R_j$ to do an alternating extraction protocol, and output $(S^j_{h1}, S^j_{h2})=\laext(\Y^j_h, R_j)$, where each $S^j_{hi}$ has $(\frac{\log^{\log a} L}{a^{j-1}} -1)d_2$ bits. Note that altogether we get $2^{j+1}$ outputs and relabel them as $\Y^{j+1}_1, \cdots, \Y^{j+1}_{2^{j+1}}$. 

\item After the previous step, we get $2 \log L$ outputs. Let them be $\Y_1, \cdots, \Y_{2 \log L}$, and output $\W_x=\nipm(\overline{V}, \X, \Y_1, \cdots, \Y_{2 \log L})$ with $m$ bits. Let $\W_y=\Ext(\Y, \W_x)$ with $d_2$ bits.
\end{enumerate}
\end{construction}

We now have the following lemma.

\BL \label{lem:nnipm3} 
There is a constant $c>1$ such that suppose we have the following random variables and conditions:
\begin{itemize}
\item $V,V'$, each supported on a boolean $L\times m$ matrix s.t for any $i \in [L]$, $V_i = U_m$. In addition, there exists an $h \in [L]$ such that $(V_h,V'_h)=(U_m,V'_h)$.
 \item $\Y, \Y'$, each supported on $n'$ bits, where $\Y$ is uniform.
 \item $\X, \X'$, each supported on $n$ bits, where $\X$ is uniform. In addition, $\X$ is independent of $(V, V')$, and $(V, V', \X, \X')$ is independent of $(\Y, \Y')$.
 \item $m \geq c \log (n'/\e)$, $n \geq 20c \log \log L \log (n'/\e)+6m$ and $n' \geq 20c \log^{\log a} L \log (n/\e)$.
 \end{itemize}
 Let $(\W_x, \W_y)$ be the outputs of $(\nipm_x, \nipm_y)$ on $(V, \X, \Y)$ and $(\W'_x, \W'_y)$ be the outputs of the $(\nipm_x, \nipm_y)$ on $(V', \X', \Y')$. Then 
 
 \[(\W_x, \W'_x, \Y, \Y') \approx_{O(L \e)} (U_{m}, \W'_x, \Y, \Y')\]
 
 and
 
  \[(\W_y, \W'_y, V, V', \X, \X') \approx_{O(L \e)} (U_{O(\log(n/\e))}, \W'_y, V, V', \X, \X').\]
\EL

\begin{proof}
First, since $(V, V', \X, \X')$ is independent of $(\Y, \Y')$, as long as $c$ is large enough, by Theorem~\ref{thm:optext} we know that for any $i \in [L]$, 
\[(\overline{\Y}_i, V) \approx_{\e} (U_d, V).\]

In addition, suppose for some $h \in [L]$ we have that $(V_h,V'_h)=(U_m,V'_h)$, then we can first fix $V'_h$ and then $\overline{\Y}_h$. Conditioned on this fixing $V_h$ is  still uniform, the average conditional min-entropy of $\Y_0$ is at least $10 \log \log L \cdot d -d > 3d$ and $V_h$ and $Y_0$ are still independent, thus by Theorem~\ref{thm:optext} we have that

\[(\overline{\Y}_h, \overline{\Y}'_h, V, V') \approx_{\e} (U_d, \overline{\Y}'_h, V, V').\]
In other words, the random variables $\{(\overline{\Y}_i, \overline{\Y}'_i)\}$ inherit the properties of $\{(V_i, V'_i)\}$. We now ignore the errors since this adds at most $L\e$ to the final error. Now we fix $(V, V')$. Note that conditioned on this fixing, the random variables $(\overline{\Y}_i, \overline{\Y}'_i)$ are deterministic functions of $(\Y_0, \Y'_0)$, and are thus independent of $(\X, \X')$. Furthermore, we have that conditioned on this fixing, $\X$ is still uniform. In addition, even conditioned on the fixing of $(\X_0, \X'_0)$, the average conditional min-entropy of $\X$ is at least $20c \log \log L \log (n'/\e)+6m-2 \cdot 10 \log \log L \cdot d_1=6m$. Thus by the same argument before we have that for any $i \in [L]$,

\[(\overline{V}_i, \Y_0, \X_0, \X'_0) \approx_{\e} (U_m, \Y_0, \X_0, \X'_0),\]

and that there exists an $h \in [L]$ such that

\[(\overline{V}_h, \overline{V}'_h, \Y_0, \Y'_0, \X_0, \X'_0) \approx_{\e} (U_m, \overline{V}'_h, \Y_0, \Y'_0, \X_0, \X'_0).\]

We will again ignore the error for now since this adds at most $L\e$ to the final error. Next, by Lemma~\ref{altext} we have that for any $0 \leq j \leq t-1$, 

\[(R_{j+1}, (R_1, R'_1, \cdots, R_j, R'_j), \Y_0, \Y'_0) \approx_{O(t\e)} (U_{4d_1}, (R_1, R'_1, \cdots, R_j, R'_j), \Y_0, \Y'_0).\]

Thus by a hybrid argument and  the triangle inequality, we have that

\[(\Y_0, \Y'_0,  R_1, R_1', \cdots, R_{t}, R'_{t}) \approx_{O(t^2 \e)} (\Y_0, \Y'_0,  U_{4d_1}, R'_1, \cdots, U_{4d_1}, R'_t),\] where each $U_{4d_1}$ is independent of all the previous random variables (but may depend on later random variables). From now on, we will proceed as if each $R_{j}$ is uniform given $(\Y_0, \Y'_0,  \{R_1, R_1', \cdots, R_{j-1}, R'_{j-1}\})$, since this only adds $O(t^2 \e)$ to the final error.

Now we can fix $(\Y_0, \Y'_0)$. Note that conditioned on this fixing, $(\overline{V}, \overline{V}', R_1, R_1', \cdots, R_{t}, R'_{t})$ are deterministic functions of $(V, V', \X, \X')$, and thus independent of $(\Y, \Y')$. Also note that conditioned on this fixing, the average conditional min-entropy of $\Y$ is at least $20 \log^{\log a} L \cdot d_2-2 \cdot 10 \log \log L \cdot d_2 > a^2 \log^{\log a} L \cdot d_2$. We now prove the following claim. 

\BCM
Let $\overline{R_j}=(R_1, \cdots, R_{j})$. Suppose that at the beginning of the $j$'th iteration, we have that conditioned on the fixing of $\overline{R_{j-1}}$, the following holds.

\begin{enumerate}
\item , $(\X, \X')$ is independent of $(\Y, \Y')$, and $(\Y_1, \Y'_1, \cdots, \Y_{2^j}, \Y'_{2^j})$ is a deterministic function of $(\Y, \Y')$.
\item For every $h \in [2^j]$, the average conditional min-entropy of $\Y_h$ given $(\Y_1, \Y'_1, \cdots, \Y_{h-1}, \Y'_{h-1})$ is at least  $(\frac{\log^{\log a} L}{a^{j-2}} -1)d_2$.
\end{enumerate}

Then at the end of the $j$'th iteration, the following holds.
\begin{enumerate}
\item Conditioned on the fixing of $\overline{R_{j}}$, $(\X, \X')$ is independent of $(\Y, \Y')$, and $(\Y_1, \Y'_1, \cdots, \Y_{2^{j+1}}, \Y'_{2^{j+1}})$ is a deterministic function of $(\Y, \Y')$.
\item For every $h \in [2^{j+1}]$, 

\[(\Y_h, (\Y_1, \Y'_1, \cdots, \Y_{h-1}, \Y'_{h-1}), \overline{R_{j}} ) \approx_{\e} (U_{(\frac{\log^{\log a} L}{a^{j-1}} -1)d_2}, (\Y_1, \Y'_1, \cdots, \Y_{h-1}, \Y'_{h-1}), \overline{R_{j}}).\]
\end{enumerate}
\ECM

\begin{proof} [Proof of the claim]
First, since the computation in the $j$'th iteration only involves $(R_j, R'_j)$ and $(\Y_1, \Y'_1, \cdots, \Y_{2^j}, \Y'_{2^j})$, and $(R_j, R'_j)$ is a deterministic function of $(\X, \X')$ conditioned on the fixing of the previous random variables, we know that at the end of the $j$'th iteration, conditioned on the fixing of $(R_1, \cdots, R_{j})$ we have that $(\X, \X')$ is independent of $(\Y, \Y')$, and $(\Y_1, \Y'_1, \cdots, \Y_{2^{j+1}}, \Y'_{2^{j+1}})$ is a deterministic function of $(\Y, \Y')$.

Next, we use $(Z_1, Z'_1, \cdots, Z_{2^{j+1}}, Z'_{2^{j+1}})$ to represent the outputs computed from $(R_j, R'_j)$ and $(\Y_1, \Y'_1, \cdots, \Y_{2^j}, \Y'_{2^j})$, and assume that $2\ell-1 \leq h \leq 2\ell$ for some $\ell$, then $Z_h$ is obtained from $\Y_{\ell}$. We can now first fix $(\Y_1, \Y'_1, \cdots, \Y_{\ell-1}, \Y'_{\ell-1})$, and conditioned on this fixing $\Y_{\ell}$ has average conditional min-entropy at least $(\frac{\log^{\log a} L}{a^{j-2}} -1)d_2$. Now by Lemma~\ref{altext} we have that

\[(S^{\ell}_1,  R_j, R'_j) \approx_{\e} (U_{(\frac{\log^{\log a} L}{a^{j-1}} -1)d_2}, R_j, R'_j)\]

and

\[(S^{\ell}_2, S^{\ell}_1, S'^{\ell}_1, R_j, R'_j) \approx_{\e} (U_{(\frac{\log^{\log a} L}{a^{j-1}} -1)d_2}, S^{\ell}_1, S'^{\ell}_1, R_j, R'_j),\]

since $(\frac{\log^{\log a} L}{a^{j-2}} -1)d_2 \geq 2 \cdot (\frac{\log^{\log a} L}{a^{j-1}} -1)d_2+(1+\alpha)(\frac{\log^{\log a} L}{a^{j-1}} -1)d_2+d_2$ and $4d_1 \geq 2d_1+1.1d_1+0.9d_1$. Thus as long as the constant $c$ is large enough one can make sure that $min\{d_2, 0.9d_1\} \geq 2\log(1/\e)$, and we can extract $(\frac{\log^{\log a} L}{a^{j-1}} -1)d_2$ bits from entropy $(1+\alpha)(\frac{\log^{\log a} L}{a^{j-1}} -1)d_2$ and $d_1$ bits from entropy $1.1 d_1$. Note that $(Z_1, Z'_1, \cdots, Z_{2\ell-2}, Z'_{2\ell-2})$ are computed from $(\Y_1, \Y'_1, \cdots, \Y_{\ell-1}, \Y'_{\ell-1})$ and $(R_j, R'_j)$, and $(\Y_1, \Y'_1, \cdots, \Y_{\ell-1}, \Y'_{\ell-1})$ are already fixed. Thus the second part of the claim also holds.
\end{proof}

Now note that at the beginning of the first iteration, the condition of the claim holds. Thus if we ignore the errors, then we can apply the claim repeatedly until the end of the iteration. At this time for each $h \in [\log L]$ we have that $\Y_h$ has at least $(\frac{\log^{\log a} L}{a^{\log \log L-1}} -1)d_2 > d_2$ bits. Furthermore

\[(\Y_h, (\Y_1, \Y'_1, \cdots, \Y_{h-1}, \Y'_{h-1}), \overline{R_t} ) \approx (U, (\Y_1, \Y'_1, \cdots, \Y_{h-1}, \Y'_{h-1}), \overline{R_t}).\]

The total error so far is  $O(L\e)+O(t^2\e)+\sum_{j=0}^{\log \log L} 2^j \cdot 2\e=O(L \e)$. Note that now conditioned on all the fixed random variables $(\X_0, \X'_0, \Y_0, \Y'_0, \overline{R_t})$ (note that $\overline{R_t}$ is a deterministic function of $(\X_0, \X'_0, \Y_0, \Y'_0)$, we have that $(V, V', \Y_1, \Y'_1, \cdots, \Y_{2 \log L}, \Y'_{2 \log L}, \X, \X')$ satisfies the conditions of the Lemma~\ref{lem:nnipm}, since the average conditional min-entropy of $X$ is at least $n-20 \log \log L \cdot d_1 \geq 6m$. Now we can apply Lemma~\ref{lem:nnipm} to show that 

\[(\W_x, \W'_x, \Y, \Y') \approx (U_{m}, \W'_x, \Y, \Y'),\]

where the total error is $O(L \e)+O(L \e)=O(L \e)$. Furthermore, note that conditioned on the fixing of $(\Y_1, \Y'_1, \cdots, \Y_{2 \log L}, \Y'_{2 \log L})$, we have that $(\W_x, \W'_x)$ is a deterministic function of $(V, V', \X, \X')$, and thus independent of $(\Y, \Y')$. Also note that $\Y$ has average conditional min-entropy at least $20c \log^{\log a} L \log (n/\e)- 4 \log L d_2 > 10d_2$. Thus by Theorem~\ref{thm:optext} we have that 

\[(\W_y, \W'_y, \W_x, \W'_x) \approx (U_{d_2}, \W'_y, \W_x, \W'_x),\]
where the error is $O(L \e)+O(\e)=O(L \e)$. Note that given $(\W_x, \W'_x)$, we have that $(\W_y, \W'_y)$ is a deterministic function of $(\Y, \Y')$. Thus we also have that
\[(\W_y, \W'_y, V, V', \X, \X') \approx_{O(L \e)} (U_{d_2}, \W'_y, V, V', \X, \X').\]
\end{proof}

\section{Correlation Breaker with Advice}\label{sec:advcb}
We now use our non-malleable independence preserving mergers to construct improved correlation breakers with advice. A correlation breaker uses independent randomness to break the correlations between several correlated random variables. The first correlation breaker appears implicitly in the author's work \cite{Li13b}, and this object is strengthened and formally defined in \cite{Cohen15}. A correlation breaker with advice additionally uses some string as an advice. This object was first introduced and used without its name in \cite{CGL15}, and then explicitly defined in \cite{Coh15nm}. 

\BD [Correlation breaker with advice] A function

\[\acb: \bits^n \times \bits^d \times \bits^L \to \bits^m\] is called a $(k, k', \eps)$-correlation breaker with advice if the following holds. Let $Y, Y'$ be $d$-bit
random variables such that $H_{\infty}(Y) \geq k'$. Let $X, X'$ be $n$-bit random variables with $H_{\infty}(X) \geq k$, such that $(X, X')$ is independent of $(Y, Y')$. Then, for any pair of distinct $L$-bit strings $\alpha, \alpha'$,

\[(\acb(X,Y,\alpha), \acb(X',Y',\alpha')) \approx_{\eps} (U,\acb(X',Y',\alpha')).\] 
In addition, we say that $\acb$ is strong if
\begin{align*}
&(\acb(X,Y,\alpha), \acb(X',Y',\alpha'), Y, Y') \\ \approx_{\eps} &(U,\acb(X',Y',\alpha'), Y, Y').
\end{align*}
\ED

Our construction needs the following flip-flop extraction scheme, which was constructed by Cohen \cite{Cohen15} using alternating extraction, based on a previous similar construction of the author \cite{Li13b}. The flip-flop function can be viewed as a basic correlation breaker, which (informally) uses an independent source $\X$ to break the correlation between two r.v's $\Y$ and $\Y'$, given an advice bit. 

\begin{theorem}[\cite{Cohen15,CGL15}]\label{flip} There exists a constant $c_{\ref{flip}}$ such that for all $n>0$ and any $\epsilon>0$, there exists an explicit function $\flip:\zo^n \times \zo^d \rightarrow \zo^m$, $m=0.4 k$,   satisfying the following: Let $\X$ be an $(n,k)$-source, and $\X'$ be a random variable on $n$ bits arbitrarily correlated with $\X$. Let $\Y$ be an independent uniform seed on $d$ bits, and $\Y'$ be a random variable on $d$ bits arbitrarily correlated with $\Y$. Suppose $(\X, \X'$) is independent of $(\Y, \Y')$.  If $k,d \ge C_{\ref{flip}}\log(n/\epsilon)$,  then for any bit $b$, 
$$|\flip(\X,\Y,b), \Y,\Y' - U_m,\Y,\Y'| \le \epsilon.$$
Furthermore, for any bits $b, b'$ with $b \neq b'$, 
\begin{align*} 
|&\flip(\X,\Y,b),\flip(\X',\Y',b'),\Y,\Y' \\ &- U_m,\flip(\X',\Y',b'),\Y,\Y'| \le \epsilon.
\end{align*}
\end{theorem}

\subsection{Asymmetric correlation breaker}

We will present correlation breakers that use general $\nipm$s. By plugging in various  $\nipm$s this gives different correlation breakers.

\begin{construction}\label{con:advcb}

Inputs:
\begin{itemize}
\item Let $\ell, m \in \N$ be two integers, $\e>0$ be an error parameter.

\item $X, Y$, two independent sources on $n$ bits and $s$ bits respectively, with min-entropy at least $n-\ell$ and $s-\ell$.

\item an advice string $\alpha \in \bits^L$.

\item An $(L, d_1, d_2, O(L\e))$-$\nipm_x$ that is strong in $Y$.

\item Let $\bip$ be the two source extractor from Theorem~\ref{thm:ip}. 
\end{itemize}

\begin{enumerate}
\item Let $d'=O(\log(max\{n, s\}/\e))$ be the seed length of the extractor from Theorem~\ref{thm:optext}, and let $d=8 d'$. Let $X^0$ be a slice of $X$ with length $d+2 \ell+2 \log(1/\e)$, and $Y^0$ be a slice of $Y$ with length $d+2 \ell+2 \log(1/\e)$.

\item Compute $Z=\bip(X^0, Y^0)$ and output $d$ bits.

\item Use $X$ and $Z$ to do an alternating extraction, and output two random variables $(X_0, X_1)=\laext(X, Z)$ where each $X_i$ has $3m$ bits.

\item Use $Y$ and $Z$ to do an alternating extraction, and output two random variables $(Y_0, Y_1)=\laext(Y, Z)$ where each $Y_i$ has $3d$ bits. 

\item Use $X_1, Y_1, \alpha$ to obtain an $L \times m$ matrix $V$, where for any $i \in [L]$, $V_i=\flip(X_1, Y_1, \alpha_i)$ and outputs $m$ bits.

\item Compute $\hat{X}=\Ext(X, Y_0)$ and output $n/2$ bits. Compute $\hat{Y}=\Ext(Y, X_0)$ and output $s/2$ bits. Here $\Ext$ is the strong seeded extractor from Theorem~\ref{thm:optext}.

\item Output $\hat{V}=\nipm_x(V, \hat{X}, \hat{Y})$. 
\end{enumerate}
\end{construction}

We now have the following lemma.

\BL \label{lem:advcb}
There exists a costant $c>1$ such that the following holds. Suppose that there exists an $(L, d_1, d_2, O(L\e))$-$\nipm$ that is strong in $Y$ which outputs $m$ bits, then there exists an explicit $(n-\ell, s-\ell, O(L \e))$ $\acb: \bits^n \times \bits^s \times \bits^L \to \bits^m$ as long as  $m \geq c \log(max\{n, s\}/\e)$, $n \geq 20m+2d_1+5 \ell+4 \log(1/\e)$ and $s \geq m+2d_2+5 \ell+4 \log(1/\e)$.
\EL

 \begin{proof}
Throughout the proof we will use letters with prime to denote the corresponding random variables obtained from $(X', Y', \alpha')$. First, notice that both $X^0$ and $Y^0$ have min-entropy at least $d+ \ell+2 \log(1/\e)$. Thus by Theorem~\ref{thm:ip} we have that 
 
 \[(Z, X^0) \approx_{\e} (U_d, X^0)\] 
 
 and 
 
  \[(Z, Y^0) \approx_{\e} (U_d, Y^0).\]
  
We now ignore the error $\e$. Note that conditioned on the fixing of $(X^0, X'^0)$, $(Z, Z')$ is a deterministic function of $(Y^0, Y'^0)$, and thus independent of $(X, X')$. Moreover, the average conditional min-entropy of $X$ given this fixing is at least $n -\ell-2(d+2 \ell+2 \log(1/\e)) \geq  10m$ as long as $c$ is large enough. Thus by Lemma~\ref{altext} (note that the extractor from $Z$ side can use seed length $d'$) we have that 
 
 \[(Y^0, Y'^0, X_0, X'_0, X_1, X'_1, Z, Z') \approx_{O(\e)} (Y^0, Y'^0, U_{3m}, X'_0, U_{d_1}, X'_1, Z, Z'),\]
 
where each $U_{3m}$ is uniform given the previous random variables, but may depend on later random variables. Similarly, note that conditioned on the fixing of $(Y^0, Y'^0)$, $(Z, Z')$ is a deterministic function of $(X^0, X'^0)$, and thus independent of $(Y, Y')$. Moreover, the average conditional min-entropy of $Y$ given this fixing is at least $s -\ell-2(d+2 \ell+2 \log(1/\e)) \geq 10d$. Thus by Lemma~\ref{altext} we have that 
 
 \[(Y_0, Y'_0, Y_1, Y'_1, Z, Z',  X^0, X'^0) \approx_{O(\e)} (U_{3d}, Y'_0, U_{d_2}, Y'_1, Z, Z',  X^0, X'^0),\]
 
where each $U_{3d}$ is uniform given the previous random variables, but may depend on later random variables. We can now fix $(X^0, X'^0, Y^0, Y'^0)$, and conditioned on this fixing, we have that $(X, X')$ and $(Y, Y')$ are still independent, $(X_0, X'_0, X_1, X'_1)$ is a deterministic function of $(X, X')$, and $(Y_0, Y'_0, Y_1, Y'_1)$ is a  deterministic function of $(Y, Y')$. Further they satisfy the look-ahead properties in the previous two equations. We will ignore the error for now since this only adds at most $O(\e)$ to the final error.

We now claim that conditioned on the fixing of $(X_0, X'_0, Y_0, Y'_0, Y_1, Y'_1)$ (and ignoring the error), the random variables $(V, V', \hat{X}, \hat{X'})$ and $(\hat{Y}, \hat{Y'})$ satisfy the conditions required by Lemma~\ref{lem:nnipm2}. To see this, note that if we fix $(Y_0, Y'_0, Y_1, Y'_1)$, then the average conditional min-entropy of $Y$ is at least $s -\ell-2(d+2 \ell+2 \log(1/\e))-2 \cdot 3d > 2s/3$ as long as $c$ is large enough. Thus by Theorem~\ref{thm:optext} we have that

\[(\hat{Y}, X_0, X'_0) \approx_{\e} (U_{s/2}, X_0, X'_0).\]

Thus conditioned on the further fixing of $(X_0, X'_0)$, we have that $(\hat{Y}, \hat{Y'})$ is a deterministic function of $(Y, Y')$, and $s/2 \geq d_2$. On the other hand, conditioned on the fixing of $(X_0, X'_0)$ and $(Y_0, Y'_0)$, we have $X_1$ is still close to uniform. Thus by Theorem~\ref{flip} we have that for any $i \in [L]$,

$$|V_i, Y_1,Y'_1 - U_m, Y_1,Y'_1| \le \e$$
and there exists $i \in [L]$ such that

$$|V_i, V'_i, Y_1,Y'_1 - U_m, V'_i, Y_1,Y'_1 | \le \e.$$
We now further fix $(Y_1,Y'_1)$. Note that conditioned on this fixing $(X, X')$ and $(Y, Y')$ are still independent. Furthermore $(V, V')$ is now a deterministic function of $(X_1, X'_1)$, and thus independent of $(Y, Y')$. Finally, note that conditioned on the fixing of $(X_0, X'_0, X_1, X'_1)$, the average conditional min-entropy of $X$ is at least $n -\ell-2(d+2 \ell+2 \log(1/\e)) - 2 \cdot 3m >  2n/3$. Thus by Theorem~\ref{thm:optext} we have that

\[(\hat{X}, Y_0, Y'_0) \approx_{\e} (U_{n/2}, Y_0, Y'_0).\]

Thus conditioned on the further fixing of $(Y_0, Y'_0)$, we have that $(\hat{X}, \hat{X'})$ is a deterministic function of $(X, X')$, and $n/2 \geq d_1$. Thus, even if conditioned on the fixing of $(X_0, X'_0, X_1, X'_1, Y_0, Y'_0, Y_1, Y'_1)$, we have that $(\hat{X}$ is close to $U_{n/2}$. Since $(V, V')$ is obtained from $(X_1, X'_1, Y_1, Y'_1)$, we know that $(\hat{X}$ is close to uniform even given $(X_0, X'_0, Y_0, Y'_0, Y_1, Y'_1)$ and $(V, V')$. Thus by Lemma~\ref{lem:nnipm2} we have that 

\[(\hat{V}, \hat{V'}, Y, Y') \approx (U_m,  \hat{V'}, Y, Y'),\]

where the error is $O(L \e)+O(L \e)+O(\e)=O(L \e)$.
\end{proof}

Next we give another correlation breaker, which recycles the randomness used.
 
\begin{construction} \label{con:advcb2}

Inputs:
\begin{itemize}
\item Let $\ell, m \in \N$ be two integers, $\e>0$ be an error parameter.

\item $X, Y$, two independent sources on $n$ bits with min-entropy at least $n-\ell$.

\item an advice string $\alpha \in \bits^L$ and an integer $2 \leq t \leq L$.

\item An $(L, d_1, d_2, O(L\e))$-$\nipm_y$ that is strong in $X$.


\item Let $\bip$ be the two source extractor from Theorem~\ref{thm:ip}. 
\end{itemize}

\begin{enumerate}
\item Let $d'=O(\log(n/\e))$ be the seed length of the extractor from Theorem~\ref{thm:optext}, and let $d=8 \frac{\log L}{\log t} d'$. Let $X^0$ be a slice of $X$ with length $d+2 \ell+2 \log(1/\e)$, and $Y^0$ be a slice of $Y$ with length $d+2 \ell+2 \log(1/\e)$.

\item Compute $Z=\bip(X^0, Y^0)$ and output $d$ bits.

\item Use $X$ and $Z$ to do an alternating extraction, and output $3\frac{\log L}{\log t}+1$ random variables $X_0, \cdots, X_{3\frac{\log L}{\log t}}$ where each $X_i$ has $d_1$ bits.

\item Use $Y$ and $Z$ to do an alternating extraction, and output two random variables $Y_0, Y_1$ where each $Y_i$ has $d_2$ bits. 

\item Use $X_0, Y_0, \alpha$ to obtain an $L \times m$ matrix $V^0$, where for any $i \in [L]$, $V^0_i=\flip(X_0, Y_0, \alpha_i)$ and outputs $m$ bits.

\item For $i=1$ to $\frac{\log L}{\log t}$ do the following. Merge every $t$ rows of $V^{i-1}$ using $\nipm_y$  and $(X_{3i-2}, Y_i)$, and output $d'$ bits. Concatenate the outputs to become another matrix $W^i$. Note that  $W^i$ has $L/t^i$ rows. Then for every row $j \in [L/t^i]$, compute $V^i_j=\Ext(X_{3i}, W^i_j)$ to obtain a new matrix $V^i$. Finally let $Y_{i+1}=\Ext(Y, X_{3i-1})$ and output $d_2$ bits.

\item Output $\hat{V}=V^{\frac{\log L}{\log t}}$. 
\end{enumerate}
\end{construction}

We now have the following lemma.

\BL \label{lem:advcb2}
There exists a costant $c>1$ such that the following holds. Suppose that for any $t \in \N$ there exists an $(t, d_1, d_2, O(t\e))$-$\nipm_y$ that is strong in $X$ which outputs $d'=O(\log(n/\e))$ bits, then there exists an explicit $(n-\ell, n-\ell, O(L \e))$ correlation breaker with advice $\acb: \bits^n \times \bits^n \times \bits^L \to \bits^m$ as long as  $d_1 \geq 4m$, $m \geq c \log (d_2/\e)$, and $n \geq c  \frac{\log L}{\log t} \log(n/\e)+max\{ 8\frac{\log L}{\log t} d_1, 2t \cdot d'+4d_2\}+5 \ell+4 \log(1/\e)$.
\EL

 \begin{proof}
Throughout the proof we will use letters with prime to denote the corresponding random variables obtained from $(X', Y', \alpha')$. First, notice that both $X^0$ and $Y^0$ have min-entropy at least $d+ \ell+2 \log(1/\e)$. Thus by Theorem~\ref{thm:ip} we have that 
 
 \[(Z, X^0) \approx_{\e} (U_d, X^0)\] 
 
 and 
 
  \[(Z, Y^0) \approx_{\e} (U_d, Y^0).\]
  
We now ignore the error $\e$. Note that conditioned on the fixing of $(X^0, X'^0)$, $(Z, Z')$ is a deterministic function of $(Y^0, Y'^0)$, and thus independent of $(X, X')$. Moreover, the average conditional min-entropy of $X$ given this fixing is at least $n -\ell-2(d+2 \ell+2 \log(1/\e)) \geq  8\frac{\log L}{\log t} d_1$ as long as $c$ is large enough. Thus by Lemma~\ref{altext} (note that the extractor from $Z$ side can use seed length $d'$) we have that 
 
 \[(Y^0, Y'^0, Z, Z', X_0, X'_0, \cdots, X_{3\frac{\log L}{\log t}}, X'_{3\frac{\log L}{\log t}}) \approx_{O((\frac{\log L}{\log t})^2 \e)} (Y^0, Y'^0, Z, Z', U_{d_1}, X'_0, \cdots, U_{d_1}, X'_{3\frac{\log L}{\log t}}),\]
 
where each $U_{d_1}$ is uniform given the previous random variables, but may depend on later random variables. Similarly, note that conditioned on the fixing of $(Y^0, Y'^0)$, $(Z, Z')$ is a deterministic function of $(X^0, X'^0)$, and thus independent of $(Y, Y')$. Moreover, the average conditional min-entropy of $Y$ given this fixing is at least $n -\ell-2(d+2 \ell+2 \log(1/\e)) \geq 4d_2$. Thus by Lemma~\ref{altext} we have that 
 
 \[(Z, Z',  X^0, X'^0, Y_0, Y'_0, Y_1, Y'_1) \approx_{O(\e)} (Z, Z',  X^0, X'^0, U_{d_2}, Y'_0, U_{d_2}),\]
 
where each $U_{d_2}$ is uniform given the previous random variables, but may depend on later random variables. We can now fix $(X^0, X'^0, Y^0, Y'^0)$, and conditioned on this fixing, we have that $(X, X')$ and $(Y, Y')$ are still independent, $(X_0, X'_0, \cdots, X_{3\frac{\log L}{\log t}}, X'_{3\frac{\log L}{\log t}})$ is a deterministic function of $(X, X')$, and $(Y_0, Y'_0, Y_1, Y'_1)$ is a  deterministic function of $(Y, Y')$. Further they satisfy the look-ahead properties in the previous two equations. We will ignore the error for now since this only adds at most $O((\frac{\log L}{\log t})^2 \e)$ to the final error.

Now by Theorem~\ref{flip} we have that for any $i \in [L]$,

$$|V^0_i, Y_0,Y'_0 - U_m, Y_0,Y'_0| \le \e$$
and there exists $i \in [L]$ such that

$$|V^0_i, V'^0_i, Y_0,Y'_0 - U_m, V'^0_i, Y_0,Y'_0 | \le \e.$$
We now further fix $(Y_0,Y'_0)$. Note that conditioned on this fixing $(X, X')$ and $(Y, Y')$ are still independent. Furthermore $(V^0, V'^0)$ is now a deterministic function of $(X_0, X'_0)$, and thus independent of $(Y, Y')$. Thus by the property of $\nipm_y$ we have that for every row $j$ in $W^1$, 

 \[(W^1_j, V^0, V'^0, X_1, X'_1) \approx_{O(t \e)} (U_{d'}, V^0, V'^0, X_1, X'_1),\]

and there exists a row $j$ such that
 \[(W^1_j, W'^1_j, V^0, V'^0, X_1, X'_1) \approx_{O(t \e)} (U_{d'}, W'^1_j, V^0, V'^0, X_1, X'_1).\]

Note that we have fixed $(X^0, X'^0, Y^0, Y'^0)$, and if we further condition on the fixing of $(X_0, X'_0, Y_0, Y'_0, X_1, X'_1)$, then $(W^1, W'^1)$ is a deterministic function of $(Y, Y')$. Furthermore $(X, X')$ and $(Y, Y')$ are still independent. We will now use induction to prove the following claim (note that we have already fixed $(X^0, X'^0, Y^0, Y'^0)$). 

\BCM
Let $T_i=(Y_0, Y'_0, X_0, X'_0, \cdots, X_{3i-2}, X'_{3i-2})$. In the $i$' th iteration, the following holds.
\begin{enumerate}
\item Conditioned on the further fixing of $T_i$, we have that $(X, X')$ and $(Y, Y')$ are still independent, and furthermore $(W^i, W'^i)$ is a deterministic function of $(Y, Y')$.

\item For every row $j$ in $W^i$,

\[(W^i_j, T_i) \approx_{\e_i} (U_{d'}, T_i),\]

and there exists a row $j$ such that

 \[(W^i_j, W'^i_j, T_i) \approx_{\e_i} (U_{d'}, W'^i_j, T_i),\]
where $\e_i=O(\sum_{j=1}^i t^j \e)$. 
\end{enumerate}
\ECM

\begin{proof}[Proof of the claim]
The base case of $i=1$ is already proved above. Now suppose the claim holds for the $i$'th iteration, we show that it also holds for the $i+1$'th iteration. 

To see this, note that conditioned on the fixing of $T_i$, $(X, X')$ and $(Y, Y')$ are still independent, and furthermore $(W^i, W'^i)$ is a deterministic function of $(Y, Y')$ and thus independent of $(X, X')$. Note that $Y_{i+1}$ is computed from $Y$ and $X_{3i-1}$ while $V^i$ is computed from $X_{3i}$ and $W^i$. Thus if we further fix $X_{3i-1}, X'_{3i-1}$ and $(W^i, W'^i)$, then $(X, X')$ and $(Y, Y')$ are still independent, and furthermore $Y_{i+1}$ is a deterministic function of $Y$ and $V^i$ is a deterministic function of $X_{3i}$. Now $W^{i+1}$ is computed from $V^i$, $X_{3i+1}$ and $Y_{i+1}$. Thus if we further fix $(X_{3i}, X'_{3i})$ and $(X_{3i+1}, X'_{3i+1})$ (i.e., we have fixed $T_{i+1}$) then $(X, X')$ and $(Y, Y')$ are still independent, and furthermore $(W^{i+1}, W'^{i+1})$ is a deterministic function of $(Y, Y')$.

Next, let $h$ be the row in $W^i$ such that 
 \[(W^i_h, W'^i_h, T_i) \approx_{\e_i} (U_{d'}, W'^i_h, T_i).\]

Note that $V^i$ has the same number of rows as $W^i$, and consider the merging of some $t$ rows in $V^i$ that contain row $h$ into $W^{i+1}_j$ (the merging of the other rows is similar and simpler). Without loss of generality assume that these  $t$ rows are row $1, 2, \cdots, t$.

First, since for every row $j$ in $W^i$,

\[(W^i_j, T_i) \approx_{\e_i} (U_{d'}, T_i),\]

and rows $h$ in $W^i$ and $W'^i$ satisfy the independence property, by Theorem~\ref{thm:optext} (and ignoring the error $\e_i$) we have that for every $j \in [t]$, 

\[(V^i_j, T_i, X_{3i-1}, X'_{3i-1}, W^i_j, W'^i_j) \approx_{\e} (U_{m}, T_i, X_{3i-1}, X'_{3i-1}, W^i_j, W'^i_j),\]

and 

\[(V^i_h, V'^i_h, T_i, X_{3i-1}, X'_{3i-1}, W^i_j, W'^i_j) \approx_{\e} (U_{m}, V'^i_h, T_i, X_{3i-1}, X'_{3i-1}, W^i_j, W'^i_j).\]

This is because $X_{3i}$ has average conditional min-entropy at least $d_1$ even conditioned on the fixing of $(X_{3i-1}, X'_{3i-1})$. We now ignore the error $\e$. Note that conditioned on the fixing of $(W^i_j, W'^i_j)$, we have that $(V^i_j, V'^i_j)$ is a deterministic function of $(X_{3i}, X'_{3i})$, and thus independent of $(Y, Y')$. We now fix $\{(W^i_j, W'^i_j), j \in [t]\}$. Note that conditioned on this fixing $\{V^i_j, j \in [t]\}$ and $\{V'^i_j, j \in [t]\}$ each is a $t \times m$ matrix, and a deterministic function of $(X_{3i}, X'_{3i})$. Further note that they form two matrices that meet the condition to apply an $\nipm$. Since $\{(W^i_j, W'^i_j), j \in [t]\}$ is a deterministic function of $(Y, Y')$, conditioned on this fixing $(X, X')$ and $(Y, Y')$ are still independent. Furthermore the average conditional min-entropy of $Y$ is at least $n -\ell-2(d+2 \ell+2 \log(1/\e))-2d_2-2t d' \geq 2d_2$. Thus by Theorem~\ref{thm:optext} we have that

\[(Y_{i+1}, X_{3i-1}) \approx_{\e} (U_{d_2}, X_{3i-1}). \]

Note that conditioned on the fixing of $X_{3i-1}$, we have that $Y_{i+1}$ is a deterministic function of $Y$. Thus we can now further fix $(X_{3i-1}, X'_{3i-1})$, and conditioned on this fixing, $Y_{i+1}$ is still close to uniform. To conclude, now conditioned on the fixing of $\{(W^i_j, W'^i_j), j \in [t]\}$ and $(X_{3i-1}, X'_{3i-1})$, we have that $\{V^i_j, j \in [t]\}$ and $\{V'^i_j, j \in [t]\}$ each is a $t \times m$ matrix, and a deterministic function of $(X_{3i}, X'_{3i})$; $Y_{i+1}$ is still close to uniform and $(Y_{i+1}, Y'_{i+1})$ is a deterministic function of $(Y, Y')$. Furthermore $X_{3i+1}$ is close to uniform. Now we can use the property of $\nipm_y$ to show that after merging these $t$ rows, the corresponding row $j$ in $W^{i+1}$ satisfies

\begin{align*}
 &(W^{i+1}_j, W'^{i+1}_j, T_i, X_{3i-1}, X'_{3i-1}, X_{3i}, X'_{3i}, X_{3i+1}, X'_{3i+1}) \\ \approx_{ t\e} &(U_{d'}, W'^{i+1}_j, T_i, X_{3i-1}, X'_{3i-1}, X_{3i}, X'_{3i}, X_{3i+1}, X'_{3i+1}).
 \end{align*}

Adding back all the errors we get that 

 \[(W^{i+1}_j, W'^{i+1}_j, T_{i+1}) \approx_{ \e_{i+1}} (U_{d'}, W'^{i+1}_j, T_{i+1}),\]

where $ \e_{i+1}=t \e_i+O(t \e)=O(\sum_{j=1}^{i+1} t^j \e)$.
\end{proof}

Now we are basically done. In the last iteration we know that $W^{\frac{\log L}{\log t}}$ has reduced to one row, and $W^{\frac{\log L}{\log t}}$ is close to uniform given $W'^{\frac{\log L}{\log t}}$. Also conditioned on the fixing of $T_{\frac{\log L}{\log t}}$ they are deterministic functions of $(Y, Y')$. Thus when we use $W^{\frac{\log L}{\log t}}$ to extract $V^{\frac{\log L}{\log t}}$ from $X_{3\frac{\log L}{\log t}}$, by Theorem~\ref{thm:optext} we have that 

\[(\hat{V}, \hat{V'}, Y, Y') \approx (U_m,  \hat{V'}, Y, Y'),\]

where the error is $O(\sum_{j=1}^{\frac{\log L}{\log t}} t^j \e)+O((\frac{\log L}{\log t})^2 \e)=O(L \e)$.
\end{proof}

%% file: ext.tex
\section{The Constructions of Non-Malleable Extractors}\label{sec:snmext}
In this section we construct our improved seeded non-malleable extractors and seedless non-malleable extractors. Both the constructions follow the general approach developed in recent works \cite{CGL15, CL16, Coh16, Li17}, i.e., first obtaining an advice and then applying an appropriate correlation breaker with advice. First we need the following advice generator from \cite{CGL15}.

\begin{theorem}[\cite{CGL15}]\label{adv_gen1} There exist a constant $c>0$ such that  for all $n>0$ and any $\epsilon>0$, there exists an explicit function $\adg:\zo^n \times \zo^d \rightarrow \zo^{L}$ with $L=c \log (n/\epsilon)$ satisfying the following: Let $X$ be an $(n,k)$-source, and $Y$ be an independent uniform seed on $d$ bits. Let $Y'$ be a random variable on $d$ bits s.t $Y' \neq Y$, and $(Y, Y')$ is independent of $X$. Then with probability at least $1-\epsilon$, $\adg(X,Y) \neq \adg(X,Y')$. Moreover, there is a deterministic function $g$ such that $\adg(X, Y)$ is computed as follows. Let $Y_1$ be a small slice of $Y$ with length $O(\log (n/\e))$, compute $Z=\Ext(X, Y_1)$ where $\Ext$ is an optimal seeded extractor from Theorem~\ref{thm:optext} which outputs $O(\log (n/\e))$ bits. Finally compute $Y_2=g(Y, Z)$ which outputs $O(\log(1/\e))$ bits and let $\adg(X, Y)=(Y_1, Y_2)$.
\end{theorem}

For two independent sources we also have the following slightly different advice generator.

\begin{theorem}[\cite{CGL15}]\label{adv_gen2} There exist constants $0< \gamma< \beta <1$ such that  for all $n>0$ and any $\epsilon \geq \e'$ for some $\e'=2^{-\Omega(n)}$, there exists an explicit function $\adg:\zo^n \times \zo^n \rightarrow \zo^{L}$ with $L=2\beta n+O(\log(1/\e))$ satisfying the following: Let $X, Y$ be two independent $(n,(1-\gamma)n)$-sources, and $(X', Y')$ be some tampered versions of $(X, Y)$, such that $(X, X')$ is independent of $(Y, Y')$. Furthermore either $X \neq X'$ or $Y \neq Y'$. Then with probability at least $1-\epsilon$, $\adg(X,Y) \neq \adg(X',Y')$. Moreover, there is a deterministic function $g$ such that $\adg(X, Y)$ is computed as follows. Let $X_1, Y_1$ be two small slice of $X, Y$ respectively, with length $\beta n$, compute $Z=\bip(X, Y_1)$ where $\bip$ is the inner product two source extractor from Theorem~\ref{thm:ip} which outputs $\Omega(n)$ bits. Finally compute $X_2=g(X, Z), Y_2=g(Y, Z)$ which both output $O(\log(1/\e))$ bits and let $\adg(X, Y)=(X_1, X_2, Y_1, Y_2)$.
\end{theorem}

By using these advice generators, the general approach of constructing seeded non-malleable extractors and seedless non-malleable extractors can be summarized in the following two theorems.

\BT \label{thm:con1}\cite{CGL15, CL16, Coh16, Li17}
There is a constant $c>1$ such that for any $n, k, d \in \N$ and $\e_1, \e_2>0$, if there is a $(k-c \log(n/\e_1), d-c \log(n/\e_1), \e_2)$ advice correlation breaker $\acb: \zo^k \times \zo^d \times \zo^{c \log(n/\e_1)} \to \zo^m$, then there exists an $(O(k), \e_1+\e_2)$ seeded non-malleable extractor $\nm: \zo^n \times \zo^d \to \zo^m$. Furthermore if $m \geq c \log(d/\e_1)$ then there exists an $(O(k), \e_1+\e_2)$ seeded non-malleable extractor $\nm: \zo^n \times \zo^{O(d)} \to \zo^{\Omega(k)}$. 
\ET
\begin{thmproof}[Sketch]
The seeded non-malleable extractor is constructed as follows. First use the seed and the source to obtain an advice as in Theorem~\ref{adv_gen1} with error $\e_1/3$, however when we compute $Z=\Ext(X, Y_1)$ we in fact output $Z_1=\Ext(X, Y_1)$ with $k$ bits and choose $Z$ to be a small slice of $Z_1$ with length $O(\log (n/\e))$. Then we can fix the random variables $(Y_1, Y'_1, Z, Z', Y_2, Y'_2)$. Note that conditioned on this fixing $(X, X')$ is still independent of $(Y, Y')$, and $(Z_1, Z'_1)$ is a deterministic function of $(X, X')$ thus is independent of $(Y, Y')$. Furthermore with probability $1-\e_1/3$, $Z_1$ has min-entropy at least $k-O(\log(n/\e_1))$ and $Y$ has min-entropy at least $d-O(\log(n/\e_1))$. We can now apply the correlation breaker to $(Z_1, Y)$ and the advice to get the desired output, where the total error is at most $\e_1/3+\e_1/3+\e_1/3+\e_2=\e_1+\e_2$. If the output $m$ is large enough (i.e., $m \geq c \log(d/\e_1)$), then we can use it to extract from $Y$ and then extract again from $Z_1$ to increase the output length to $\Omega(k)$.
\end{thmproof}

\BT \label{thm:con2}\cite{CGL15, CL16, Coh16, Li17}
There are constants $c>1$, $0< \gamma< \beta <1/100$ such that for any $n \in \N$ and $\e_1, \e_2>0$, if there is a $((1-2\beta) n-c \log(n/\e_1), (1-2\beta) n-c \log(n/\e_1), \e_2)$ advice correlation breaker $\acb: \zo^{n} \times \zo^n \times \zo^{2\beta n+c\log(1/\e_1)} \to \zo^m$, then there exists an $((1-\gamma)n, (1-\gamma)n, \e_1+\e_2)$ non-malleable two source extractor $\nm: \zo^n \times \zo^n \to \zo^m$. Furthermore if $m \geq c \log(n/\e_1)$ then there exists an $((1-\gamma)n, (1-\gamma)n, \e_1+\e_2)$ non-malleable two source extractor $\nm: \zo^n \times \zo^n \to \zo^{\Omega(n)}$. 
\ET

\begin{thmproof}[Sketch]
The non-malleable two-source extractor is constructed as follows. First use the two independent sources $(X, Y)$ to obtain an advice as in Theorem~\ref{adv_gen2} with error $\e_1/3$, then we can fix the random variables $(X_1, X'_1, Y_1, Y'_1, X_2, X'_2, Y_2, Y'_2)$. Note that conditioned on this fixing $(X, X')$ is still independent of $(Y, Y')$, furthermore with probability $1-\e_1/3$, both $X$ and $Y$ have min-entropy at least $(1-\gamma) n-\beta n-c\log(1/\e_1) \geq (1-2\beta)n-c\log(1/\e_1)$. We can now apply the correlation breaker to $(X, Y)$ and the advice to get the desired output, where the total error is at most $\e_1/3+\e_1/3+\e_1/3+\e_2=\e_1+\e_2$. If the output $m$ is large enough (i.e., $m \geq c \log(d/\e_1)$), then we can use it to extract from $Y$ and then extract again from $X$ to increase the output length to $\Omega(n)$. 
\end{thmproof}

Combined with our new correlation breakers with advice, we have the following new constructions of non-malleable extractors.

\BT \label{thm:snmext1}
There exists a constant $C>1$ such that for any constant $a \in \N, a \geq 2$, any $n, k \in \N$ and any $0<\e<1$ with $k \geq C(\log n+a \log(1/\e))$, there is an explicit construction of a strong seeded $(k, \e)$ non-malleable extractor $\zo^n \times \zo^d \to \zo^m$ with $d=O(\log n)+\log(1/\e)2^{O(a(\log \log (1/\e))^\frac{1}{a})}$ and $m =\Omega(k)$. Alternatively, we can also achieve entropy $k \geq C \log n+\log(1/\e)2^{C \cdot a(\log \log (1/\e))^\frac{1}{a}}$ and $d =O(\log n+a \log(1/\e))$.
\ET

\begin{thmproof}
The theorem is obtained by combining Theorem~\ref{thm:con1}, Lemma~\ref{lem:advcb} and Lemma~\ref{lem:nnipm2}. We choose an error $\e'$ to be the error in Theorem~\ref{thm:con1}, Lemma~\ref{lem:advcb} and Lemma~\ref{lem:nnipm2}. Thus the total error is $O(L \e')$ where $L=O(\log(n/\e'))$. To ensure $O(L \e')=\e$ it suffices to take $\e'=\frac{\e}{c \log(n/\e)}$ for some constant $c>1$. We know $\ell=O(\log(n/\e'))$. Therefore to apply Lemma~\ref{lem:advcb} and Lemma~\ref{lem:nnipm2}, we need to find $m, d', d_1, d_2$ such that 

\[d' \geq c (\log \max\{d_1, d_2\}+\log(1/\e')), m \geq d', d_1 \geq 8a \cdot d'+6m \text{ and } d_2 \geq 8a \cdot d'+c^{a \cdot \log ^{\frac{1}{a}} L} \cdot d'.\]

Then we can take 

\[k=O(d_1+m+\ell+\log(1/\e')) \text{ and } d=O(d_2+m+\ell+\log(1/\e')).\]


It can be seen that we can take $m=O(\log(n/\e'))$, $d'=O(\log \log n+\log(1/\e'))$, $d_1=8a \cdot d'+6m=O( \log n+a \log (1/\e'))$ and $d_2=2^{O(a(\log \log (n/\e'))^\frac{1}{a})} \cdot d' $. We now consider two cases. First, $\log(1/\e') >\frac{\log n}{c'^{a(\log \log n)^\frac{1}{a}}}$ for some large constant $c'$. In this case we have that 

\[\log(1/\e') >\frac{\log n}{c'^{a(\log \log n)^\frac{1}{a}}} > \sqrt{\log n}\]

for any $a \geq 2$. Thus 

\[\log \log (n/\e'))=\log(\log n+\log(1/\e')) < \log (\log^2(1/\e')+\log(1/\e')) < 2 \log \log (1/\e')+1.\]

Also note that $d'=O(\log \log n+\log(1/\e'))=O(\log(1/\e'))$. Thus in this case we have $d_2 \leq O(\log(1/\e'))2^{O(a(\log \log (1/\e'))^\frac{1}{a})}=\log(1/\e')2^{O(a(\log \log (1/\e'))^\frac{1}{a})}$. Next, consider the case where $\log(1/\e')  \leq \frac{\log n}{c'^{a(\log \log n)^\frac{1}{a}}}$. In this case note that we have $\log(1/\e') < \log n$ and thus $2^{O(a(\log \log (n/\e'))^\frac{1}{a})} < 2^{O(a(\log \log (n))^\frac{1}{a})}$. Therefore when $c'$ is large enough and $a \geq 2$ we have that 

\[d_2 \leq 2^{O(a(\log \log (n))^\frac{1}{a})}(\log \log n+\log(1/\e')) \leq \log n.\] 

Therefore altogether we have that $d_2 \leq (\log n+\log(1/\e')2^{O(a(\log \log (1/\e'))^\frac{1}{a})})$ and $d=O(d_2+m+\ell+\log(1/\e'))=O(\log n)+\log(1/\e')2^{O(a(\log \log (1/\e'))^\frac{1}{a})}$. Note that $\log(1/\e')=\log(1/\e)+\log (\log n+\log(1/\e))+O(1)$, a careful analysis similar as above shows that we also have that 

\[d=O(\log n)+\log(1/\e)2^{O(a(\log \log (1/\e))^\frac{1}{a})}.\]

Note that the correlation breaker is completely symmetric to both sources, and the only difference is in generating the advice. Thus after advice generation which costs both sources $O(\log(n/\e))$ entropy, we can switch the role of the seed and the source. Therefore we can also get the other setting of parameters where $k \geq C \log n+\log(1/\e)2^{C \cdot a(\log \log (1/\e))^\frac{1}{a}}$ and $d =O(\log n+a \log(1/\e))$.
\end{thmproof}

By using this theorem, we can actually improve the entropy requirement of the non-malleable extractor. Specifically, we have the following theorem.

\BT \label{thm:snmext2}
There exists a constant $C>1$ such that for any constant $a \in \N, a \geq 2$, any $n, k \in \N$ and any $0<\e<1$ with $k \geq C(\log \log n+a \log(1/\e))$, there is an explicit construction of a strong seeded $(k, \e)$ non-malleable extractor $\zo^n \times \zo^d \to \zo^m$ with $d=O(\log n)+\log(1/\e)2^{O(a(\log \log (1/\e))^\frac{1}{a})}$ and $m =\Omega(k)$. Alternatively, we can also achieve entropy $k \geq C \log \log n+\log(1/\e)2^{C \cdot a(\log \log (1/\e))^\frac{1}{a}}$ and $d =O(\log n+a \log(1/\e))$.
\ET

\begin{thmproof}
We start by taking a slice of the seed $Y_1$ with length $O(\log(n/\e))$ to extract from the source, and output some $k' =0.9k$ uniform bits with error $\e/2$. Note that conditioned on the fixing of $(Y_1, Y'_1)$ where $Y'_1$ is the tampered version, the two sources are still independent, and the seed now has average conditional entropy at least $d-O(\log(n/\e))$. We now switch the role of the seed and the source, and use the output of the extractor from the source as the seed of a non-malleable extractor and apply Theorem~\ref{thm:snmext1} with error $\e/2$, so that the final error is $\e$. 

Note that now we know the original seed is different from its tampered version, so we only need to obtain advice from the original seed and thus the advice size is $O(\log(d/\e))$. Now we only need 

\[k \geq C (\log d+a \log(1/\e))\]

and 

\[d- O(\log(n/\e)) \geq C\log k+\log(1/\e)2^{C \cdot a(\log \log (1/\e))^\frac{1}{a}}.\]

Thus we can choose 
\[k \geq C'(\log \log n+a \log(1/\e))\]
for some slightly larger constant $C'>1$, while the requirement of the seed is still

\[d=O(\log n)+\log(1/\e)2^{O(a(\log \log (1/\e))^\frac{1}{a})}.\]

Similarly, we can switch the role of the seed and the source to get the other setting of parameters. 
\end{thmproof}
The next theorem improves the seed length, at the price of using a slightly larger entropy.

\BT \label{thm:snmext3}
There exists a constant $C>1$ such that for any $n, k \in \N$ and $0<\e<1$ with $k \geq C(\log n+\log(1/\e)\log \log \log(1/\e))$, there is an explicit construction of a strong seeded $(k, \e)$ non-malleable extractor $\zo^n \times \zo^d \to \zo^m$ with $d=O(\log n+\log(1/\e)(\log \log(1/\e))^2)$ and $m =\Omega(k)$.
\ET

\begin{thmproof}
The theorem is obtained by combining Theorem~\ref{thm:con1}, Lemma~\ref{lem:advcb} and Lemma~\ref{lem:nnipm3}. Again, We choose an error $\e'$ to be the error in Theorem~\ref{thm:con1}, Lemma~\ref{lem:advcb} and Lemma~\ref{lem:nnipm3}. Thus the total error is $O(L \e')$ where $L=O(\log(n/\e'))$. To ensure $O(L \e')=\e$ it suffices to take $\e'=\frac{\e}{c \log(n/\e)}$ for some constant $c>1$. We also know $\ell=O(\log(n/\e'))$ in Lemma~\ref{lem:advcb}. Thus to apply Lemma~\ref{lem:nnipm3}, we need to find $m, d_1, d_2$ such that (for simplicity, we choose $a=4$ in Lemma~\ref{lem:nnipm3}),

\[m \geq c \log (d_2/\e'), d_1 \geq 20c \log \log L \log (d_2/\e')+6m \text{ and } d_2 \geq 20c \log^2 L \log (d_1/\e').\]

Then we can take 

\[k=O(d_1+m+\ell+\log(1/\e')) \text{ and } d=O(d_2+m+\ell+\log(1/\e')).\]
A careful but tedious calculation shows that we can choose $k \geq C(\log n+\log(1/\e')\log \log \log(1/\e'))$ for some large enough constant $C>1$, and $d=O(\log n+\log(1/\e')(\log \log(1/\e'))^2)$. Note that we can choose $m =O(\log(n/\e'))$ for a large enough constant in $O(.)$, thus by Theorem~\ref{thm:con1} we can get an output length of $\Omega(k)$. Finally, note that $\log(n/\e')=O(\log(n/\e))$, thus the theorem follows.
\end{thmproof}

Similar to what we have done above, we can also use this to get improved parameters. Specifically, we have

\BT \label{thm:snmext4}
There exists a constant $C>1$ such that for any $n, k \in \N$ and $0<\e<1$ with $k \geq C(\log \log n+\log(1/\e)\log \log \log(1/\e))$, there is an explicit construction of a strong seeded $(k, \e)$ non-malleable extractor $\zo^n \times \zo^d \to \zo^m$ with $d=O(\log n+\log(1/\e)(\log \log(1/\e))^2)$ and $m =\Omega(k)$. Alternatively, we can also achieve entropy $k \geq C(\log \log n+\log(1/\e)(\log \log(1/\e))^2)$ and seed length $d=O(\log n+\log(1/\e)\log \log \log(1/\e))$.
\ET

For non-malleable two-source extractors we have the following theorem.

\BT \label{thm:tnmext}
There exists a constant $0< \gamma< 1$ and a non-malleable two-source extractor for $(n, (1-\gamma)n)$ sources with error $2^{-\Omega(n \log \log n/\log n)}$ and output length $\Omega(n)$. 
\ET

\begin{thmproof}
The theorem is obtained by combining Theorem~\ref{thm:con2}, Lemma~\ref{lem:advcb2} and Lemma~\ref{lem:nnipm2}. Again, we choose an error $\e'$ to be the error in Theorem~\ref{thm:con1}, Lemma~\ref{lem:advcb} and Lemma~\ref{lem:nnipm3}. Thus the total error is $O(L \e')$ where $L=O(n)$. To ensure $O(L \e')=\e$ it suffices to take $\e'=\frac{\e}{c n}$ for some constant $c$. We also know $\ell=2 \beta n+o(n)$ for some constant $\beta<1/100$ in Lemma~\ref{lem:advcb2}. We choose $a=2$ in Lemma~\ref{lem:nnipm2} and thus we obtain a correlation breaker with $m=O(\log(n/\e'))$, $d_1=O(\log(n/\e'))$ and $d_2=\log (n/\e')2^{O(\sqrt{\log t})}$ where $t$ is the parameter in Construction~\ref{con:advcb2} with $t \leq L$. Note that this also satisfies that $d_1 \geq 4m$ and $m \geq c \log (d_2/\e)$ as required by Lemma~\ref{lem:advcb2}.

Now we need to ensure that

\[(1-\beta)n \geq c  \frac{\log L}{\log t} \log(n/\e')+max\{ 8\frac{\log L}{\log t} d_1, 2t \cdot d'+4d_2\}+5 \ell+4 \log(1/\e'),\]

where $d'=O(\log(n/\e'))$. We choose $t =\frac{\log L}{\log \log L}$ and this gives us

\[(1-12\beta)n \geq C \frac{\log L}{\log \log L} \log(n/\e'),\]

for some constant $C>1$. Note that $\log(n/\e')=O(\log(n/\e))$ thus we can set $\e=2^{-\Omega(n \log \log n/\log n)}$ and satisfy the above inequality.
\end{thmproof}

For applications in two-source extractors, we first need the following generalization of non-malleable extractors, which allows multiple tampering.

\begin{definition}[Seeded $t$-Non-malleable extractor] A function $\snmExt:\{0,1\}^n \times \{ 0,1\}^d \rightarrow \{ 0,1\}^m$ is a seeded $t$-non-malleable extractor for min-entropy $k$ and error $\epsilon$ if the following holds : If $X$ is a  source on  $\{0,1\}^n$ with min-entropy $k$ and $\A_1, \cdots, \A_t : \{0,1\}^d \rightarrow \{0,1\}^d $ are $t$ arbitrary tampering functions with no fixed points, then
$$  \left |\snmExt(X,U_d) \scirc \{\snmExt(X,\A_i(U_d)), i \in [t]\} \scirc U_d- U_m \scirc  \{\snmExt(X,\A_i(U_d)), i \in [t]\} \scirc U_d \right | <\epsilon $$where $U_m$ is independent of $U_d$ and $X$.
\end{definition}

The following theorem is a special case of Theorem 8.6 proved in \cite{Li17}.

\BT \label{thm:nmconvert}
Suppose there is a function $f$, a constant $\gamma>0$ and an explicit non-malleable two-source extractor for $(f(\e), (1-\gamma)f(\e))$ sources with error $\e$ and output length $\Omega(f(\e))$. Then there is a constant $C>0$ such that for any $0< \e < 1$ with $k \geq C t^2(\log n +f(\e))$, there is an explicit strong seeded $t$-non-malleable extractor for  $(n, k)$ sources with seed length $d=Ct^2(\log n +f(\e))$, error $O(t \e)$ and output length $\Omega(f(\e))$. 
\ET

Combined with Theorem~\ref{thm:tnmext}, this immediately gives the following theorem.

\BT \label{thm:snmext5}
There is a constant $C>0$ such that for any $0< \e < 1$ and $n, k \in \N$ with $k \geq C t^2(\log n +\frac{\log(1/\e) \log \log (1/\e)}{\log \log \log(1/\e)})$, there is an explicit strong seeded $t$-non-malleable extractor for  $(n, k)$ sources with seed length $d=Ct^2(\log n +\frac{\log(1/\e) \log \log (1/\e)}{\log \log \log(1/\e)})$, error $O(t \e)$ and output length $\Omega(k/t^2)$. As a special case, there exists a seeded non-malleable extractor for entropy $k \geq C (\log n +\frac{\log(1/\e) \log \log (1/\e)}{\log \log \log(1/\e)})$ and seed length $d = C (\log n +\frac{\log(1/\e) \log \log (1/\e)}{\log \log \log(1/\e)})$.
\ET

Similar techniques as above can reduce the $\log n$ term in the entropy requirement to $\log \log n$, so we get

\BT \label{thm:snmext6}
There is a constant $C>0$ such that for any $0< \e < 1$ and $n, k \in \N$ with $k \geq C (\log \log n +\frac{\log(1/\e) \log \log (1/\e)}{\log \log \log(1/\e)})$, there is an explicit strong seeded non-malleable extractor for  $(n, k)$ sources with seed length and seed length $d = C (\log n +\frac{\log(1/\e) \log \log (1/\e)}{\log \log \log(1/\e)})$.
\ET

Ben-Aroya et. al \cite{BDT16} proved the following theorem.

\BT \cite{BDT16} \label{thm:gext}
Suppose there is a function $f$ and an explicit strong seeded $t$-non-malleable extractor $(n, k')$ sources with seed length and entropy requirement $d=k'=f(t, \e)$, then for every constant $\e>0$ there exist constants $t=t(\e), c=c(\e)$ and an explicit extractor $\Ext: (\zo^n)^2 \to \zo$ for two independent $(n, k)$ sources with $k \geq f(t, 1/n^c)$ and error $\e$.
\ET

Combined with Theorem~\ref{thm:snmext3}, this immediately gives the following theorem.

\BT \label{thm:text}
For every constant $\e >0$, there exists a constant $C>1$ and an explicit two source extractor $\Ext: (\zo^n)^2 \to \zo$ for entropy $k \geq C\frac{\log n \log \log n}{\log \log \log n}$ with error $\e$. 
\ET

%% file: newcode.tex
\section{Non-Malleable Two-Source Extractor and Non-Malleable Code}\label{sec:nmtext}
Formally, non-malleable codes are defined as follows. 

\BD \cite{ADKO15}
Let $\snm_k$ denote the set of trivial manipulation functions on $k$-bit strings, which consists of the identity function $I(x)=x$ and all constant functions $f_c(x)=c$, where $c \in \bits^k$. Let $E: \bits^k \to \bits^m$ be an efficient randomized \emph{encoding} function, and $D: \bits^m \to \bits^k$ be an efficient deterministic \emph{decoding} function. Let ${\mathcal F}: \bits^m \to \bits^m$ be some class of functions. We say that the pair $(E, D)$ defines an $({\mathcal F}, k, \e)$-\emph{non-malleable code}, if for all $f \in {\mathcal F}$ there exists a probability distribution $G$ over $\snm_k$, such that for all $x \in \bits^k$, we have

\[\left |D(f(E(x)))-G(x) \right | \leq \e.\]  
\ED

\begin{remark}
The above definition is slightly different form the original definition in \cite{DPW10}. However, \cite{ADKO15} shows that the two definitions are equivalent.
\end{remark}

We will mainly be focusing on the following family of tampering functions in this paper.

\BD Given any $t>1$, let ${\mathcal S}^t_n$ denote the tampering family in the $t$-\emph{split-state-model}, where the adversary applies $t$ arbitrarily correlated functions $h_1, \cdots, h_{t}$ to $t$ separate, $n$-bit parts of string. Each $h_i$ can only be applied to the $i$-th part individually.
\ED

We remark that even though the functions $h_1, \cdots, h_{t}$ can be correlated, their correlation is independent of the original codewords. Thus, they are actually a convex combination of independent functions, applied to each part of the codeword. Therefore, without loss of generality we can assume that each $h_i$ is a deterministic function, which acts on the $i$-th part of the codeword individually.We will mainly consider the case of $t=2$, i.e., the two-split-state model. We recall the original definition of non-malleable two-source extractors by Cheraghchi and Gursuswami \cite{CG14b}. First we define the following function.

\[
 \cpy(x,y) =
  \begin{cases}
   x & \text{if } x \neq \same \\
   y       & \text{if } x  = \same

  \end{cases}
\]

\begin{definition}[Seedless Non-Malleable $2$-Source Extractor]\label{def:gt2}
A function $\nmExt : (\{ 0,1\}^{n})^2 \rightarrow \{ 0,1\}^m$ is a $(k, \e)$-seedless non-malleable extractor for two independent sources, if it satisfies the following property: Let $X, Y$ be two independent $(n, k)$ sources, and $f_1, f_2 : \zo^n \to \zo^n$ be two arbitrary tampering functions, then  
\begin{enumerate}
\item $|\nmExt(X, Y)-U_m| \leq \e$.
\item There is a distribution $\mathcal D$ over $\zo^m \cup \{\same\}$ such that for an independent $Z$ sampled from $D$, we have
$$ (\nmExt(X, Y), \nmExt(f_1(X), f_2(Y))) \approx_{\e}  (\nmExt(X, Y), \cpy(Z, \nmExt(X, Y))).$$ 
\end{enumerate}
\end{definition}

Cheraghchi and Gursuswami \cite{CG14b} showed that the relaxed definition~\ref{def:t2} implies the above general definition with a small loss in parameters. Specifically, we have

\BL [\cite{CG14b}] \label{lem:nmeq} 
Let $\nmExt$ be a $(k-\log(1/\e), \e)$-non-malleable two-source extractor according to Definition~\ref{def:t2}. Then $\nmExt$ is a $(k, 4\e)$-non-malleable two-source extractor according to Definition~\ref{def:gt2}.
\EL

The following theorem was proved by Cheraghchi and Gursuswami \cite{CG14b}, which establishes a connection between seedless non-malleable extractors and non-malleable codes.

\BT \label{connection} Let $\nmExt:\{0,1\}^{n} \times \{ 0,1\}^n \rightarrow \{0,1\}^{m}$  be a polynomial time computable seedless $2$-non-malleable extractor  at min-entropy $n$ with error $\epsilon$. Then there exists an explicit non-malleable code with an efficient decoder in the $2$-split-state model with block length $=2n$, rate  $= \frac{m}{2n}$ and error $=2^{m+1}\epsilon$.
\ET

One can construct a non-malleable code in the $2$-split-state model from a non-malleable two-source extractor as follows: Given any message $s \in \{ 0,1\}^m$, the encoding $\Enc(s)$ is done by outputting a uniformly random string from the set $\nmExt^{-1}(s) \subset \{ 0,1\}^{2n}$. Given any codeword $c \in \{0,1\}^{2n}$, the decoding $\Dec(c)$ is done by outputting $\nmExt(c)$. Thus, to get an efficient encoder we need a way to efficiently uniformly sample from the pre-image of any output of the extractor. 

Since our new non-malleable two-source extractor follows the same structure as in \cite{Li17}, we can use the same sampling procedure there to efficiently uniformly sample from the pre-image of any output of the extractor. We briefly recall the construction and sampling procedure in \cite{Li17}.

\paragraph{The extractor construction and sampling.} The high level structure of the non-malleable two-source extractor in \cite{Li17} is as follows. First take two small slices $(X_1, Y_1)$ of both sources and apply the inner product based two-source extractor, as in Theorem~\ref{thm:ip}. Then, use the output to sample $O(\log(1/\e))$ bits from the encodings of both sources, using a randomness efficient sampler and an asymptotically good linear encoding of the sources. We need an asymptotically good encoding since then we only need to sample $O(\log(1/\e))$ bits to ensure that the sampling of two different codewords are different with probability at least $1-\e$. The advice is then obtained by combining the slices and the sample bits. Now, take two larger slices $(X_2, Y_2)$ of both sources and apply the correlation breaker. Finally, take another larger slice of either source (say $X_3$ from $X$) and apply a strong linear seeded extractor, which is easy to invert and has the same pre-image size for any output. By limiting the size of each slice to be small, the construction ensures that there are at least $n/2$ bits of each source that are only used in the encoding of the sources but never used in the subsequent extraction.

Now to sample uniformly from the pre-image of any output, we first uniformly independently generate the slices $(X_1, Y_1, X_2, Y_2)$ and the sampled bits $Z$. From these we can compute the coordinates of the sampled bits and the output of the correlation breaker. Now we can invert the linear seeded extractor and uniformly sample $X_3$ given the output of the extractor and the output of the correlation breaker (which is used as the seed of the linear seeded extractor). Now, to sample the rest of the bits, we need to condition on the event that the sample bits from the encoding of the sources are indeed $Z$. Note that $Z$ has size at most $\alpha n$ for some small constant $\alpha< 1/2$ since we can restrict the error to be at least some $2^{-\Omega(n)}$. Also note that for each source we have already sampled some bits but there are still at least $n/2$ un-sampled free bits, thus we insist on that no matter which $\alpha n$ columns of the generating matrix of the encoding we look at, the sub matrix corresponding to these columns and the last $n/2$ rows have full column rank. If this is true then no matter which coordinates we use and what $Z$ is, the pre-image always have the same size and we can uniformly sample from the pre-image by solving a system of linear equations.

In \cite{Li17}, we use the Reed-Solomn encoding for each source with field $\F_q$ for $q \approx n$. This is asymptotically good and also satisfies the property that any sub matrix with less columns than rows has full column rank since it is a Vandermonde matrix. However in this case each symbol has roughly $\log n$ bits so we can sample at most $n/\log n$ symbols (otherwise fixing them may already cost us all the entropy), thus the best error we can get using this encoding is $2^{-n/\log n}$. 

We now give a new construction of non-malleable two-source extractors for two $(n, (1-\gamma)n)$ sources, where $0<\gamma<1$ is some constant. First, we need the following ingredients. 

\BT [\cite{Li17}] \label{thm:iext}
There exists a constant $0<\alpha<1$ such that for any $n \in \N$ and $2^{-\alpha n}< \e<1 $ there exists a linear seeded strong extractor $\iext: \bits^n \times \bits^d \to \bits^{0.3 d}$ with $d=O(\log(n/\e))$ and the following property. If $X$ is a $(n,0.9n)$ source and $R$ is an independent uniform seed on $\{ 0,1\}^{d}$, then $$ |(\iext(X,R),R) - (U_{0.3 d},R)| \leq \e.$$ 
Furthermore for any $s \in \{ 0,1\}^{0.3 d}$ and any $r \in  \{ 0,1\}^{d}$, $| \iext(\cdot,r)^{-1}(s)|= 2^{n-0.3 d}$.
\ET


\BD[Averaging sampler \cite{Vadhan04}]\label{def:samp} A function $\samp: \{0,1\}^{r} \rightarrow [n]^{t}$ is a $(\mu,\theta,\gamma)$ averaging sampler if for every function $f:[n] \rightarrow [0,1]$ with average value $\frac{1}{n}\sum_{i}f(i) \ge \mu$, it holds that 
$$ \Pr_{i_1,\ldots,i_t \leftarrow \samp(U_{R})}\left [  \frac{1}{t}\sum_{i}f(i) < \mu - \theta \right ] \leq \gamma.$$
$\samp$ has distinct samples if for every $x \in \{ 0,1\}^{r}$, the samples produced by $\samp(x)$ are all distinct.
\ED

\BT [\cite{Vadhan04}] \label{thm:samp} Let $1 \geq \delta \geq 3\tau > 0$. Suppose that $\samp: \zo^r \to [n]^t$ is an $(\mu,\theta,\gamma)$ averaging sampler with distinct samples for $\mu=(\delta-2\tau)/\log(1/\tau)$ and $\theta=\tau/\log(1/\tau)$. Then for every $\delta n$-source $X$ on $\zo^n$, the random variable $(U_r,  X_{Samp(U_r)})$ is $(\gamma+2^{-\Omega(\tau n)})$-close to $(U_r, W)$ where for every $a \in \zo^r$, the random variable $W|_{U_r=a}$ is $(\delta-3\tau)t$-source.
\ET

\BT[\cite{Vadhan04}]\label{thm:sampler} For every $0< \theta< \mu<1$, $\gamma>0$, and $n \in \N$, there is an explicit $(\mu,\theta,\gamma)$ averaging sampler $\samp: \zo^r \to [n]^t$
 that uses
 \begin{itemize}
 \item $t$ distinct samples for any $t \in [t_0, n]$, where $t_0=O(\frac{1}{\theta^2} \log(1/\gamma))$, and
 \item $r=\log (n/t)+\log(1/\gamma)\poly(1/\theta)$ random bits.
 \end{itemize}
\ET

\subsection{A new advice generator}
Here we show that we can give a new advice generator with optimal advice length. We have the following construction. Let $(X, Y)$ be two independent $(n, (1-\tau)n)$ sources. Let $\bip$ be the inner product two-source extractor from Theorem~\ref{thm:ip}, and $\samp:$ be the sampler from Theorem~\ref{thm:samp}. Let $L>0$ be a parameter, and $c>0$ be a constant to be chosen later. We have the following algorithm.

\begin{enumerate}
\item Let $n_1=3 \tau n$. Divide $X$ into $X=(X_1, X_2)$ such that $X_1$ has $n_1$ bits and $X_2$ has $n_2=(1-3 \tau)n$ bits. Similarly divide $Y$ into $Y=(Y_1, Y_2)$ such that $Y_1$ has $n_1$ bits and $Y_2$ has $n_2=(1-3 \tau)n$ bits.

\item Compute $Z=\bip(X_1, Y_1)$ which outputs $r=\Omega(n) \leq \tau n$ bits. 

item Let $\F$ be the finite field $\F_{2^{\log n}}$. Let $n_0 = \frac{n_2}{\log n}$. Let $\RS: \F^{n_0} \rightarrow \F^{n}$ be the Reed-Solomon code encoding $n_0$ symbols of $\F$ to $n$ symbols in  $\F$ (we  slightly abuse the use of $\RS$ to denote both the code and the encoder). Thus $\RS$ is a $[n,n_0,n-n_0+1]_{n}$ error correcting code. Let $\hat{X}_2$ be $X_2$ written backwards, and similarly $\hat{Y}_2$ be $Y_2$ written backwards. Let $\overline{X}_2=\RS(\hat{X}_2)$ and $\overline{Y}_2=\RS(\hat{Y}_2)$.

\item Use $Z$ to sample $r/\log n$ distinct symbols from $\overline{X}_2$ (i.e., use each $\log n$ bits to sample a symbol), and write the symbols as a binary string $\tilde{X}_2$. Note that $\tilde{X}_2$ has $r$ bits. Similarly, use $Z$ to sample $r/\log n$ distinct symbols from $\overline{Y}_2$ and obtain a binary string $\tilde{Y}_2$ with $r$ bits.

\item Let $V_1=X_1 \circ Y_1 \circ \tilde{X}_2 \circ \tilde{Y}_2$. 

\item Take a slice of $X_2$ with length $15 \tau n$, and let it be $X_3$. Similarly, take a slice of $Y_2$ with length $10 \tau n$, and let it be $Y_3$. Compute $W=\bip(X_3, Y_3)$ which outputs $r=\Omega(n) \leq \tau n$ bits.

\item Take a slice of $X_2$ with length $40 \tau n$, and let it be $X_4$. Use $W$ and $X_4$ to do an alternating extraction protocol for $L = \log^* n$\footnote{Here by $\log^* n$ we mean the number of steps it takes to get down to $c'$ by computing $n \to c \log n$ for some constants $c, c'$.} rounds, and output $(R_1, \cdots, R_L)=\laext(X_4, W)$, where each $S_i, R_i$ used in the alternating extraction has $\tau n/ \log n$ bits.  

\item Set $i=1$ and let $n_1$ be the length of $V_1$, which is at most $8 \tau n$. While $L < c \log n_i$ do the following: encode $V_i$ to $\tilde{V_i}$ using an asymptotically good binary error correcting code. Cut $R_i$ into $O(\log n_i)$ bits. Use the sampler from Theorem~\ref{thm:sampler} and $R_i$ to sample $\log n_i$ bits of $\tilde{V_i}$, let the sampled string be $\overline{V_i}$. Set $V_{i+1}=R_i \circ \overline{V_i}$ and let $i=i+1$.

\item Finally, cut $R_i$ into $O(\log n_i)$ bits. Use the sampler from Theorem~\ref{thm:sampler} and $R_i$ to sample $L-|R_i|$ bits of $\tilde{V_i}$, let the sampled string be $\overline{V_i}$. Set $\tilde{\alpha}=R_i \circ \overline{V_i}$ which has length $L$.
\end{enumerate}

We have the following lemma.

\BL \label{lem:newadv}
There are constants $0< \tau, \mu<1$ and $C>1$ such that the following holds. Let $(X, Y)$ be two independent $(n, (1-\tau)n)$ sources, and $(X', Y')$ be their tampered versions. Assume that either the tampering function $f$ on $X$ or the tampering function $g$ on $Y$ has no fixed point. For any $L$ such that $C \leq L \leq \frac{\mu n}{\log n}$, with probability $1-2^{-\Omega(L)}$ over the fixing of $(X_1, Y_1, \tilde{X}_2, \tilde{Y}_2, X_3, Y_3, X_4)$ and the tampered versions $(X'_1, Y'_1, \tilde{X}_2', \tilde{Y}_2', X'_3, Y'_3, X'_4)$, we have that $\tilde{\alpha} \neq \tilde{\alpha}'$. Moreover, conditioned on these fixings, $X$ and $Y$ are independent, and the average conditional min-entropy of both $X$ and $Y$ is $(1-O(\tau)) n$.
\EL

\begin{proof}
As usual we use letters with primes to denote the tampered versions of random variables. First note that both $X_1$ and $Y_1$ have min-entropy at least $2 \tau n$, thus by Theorem~\ref{thm:ip}, we have that

\[(Z, X_1) \approx_{2^{-\Omega(n)}} (U, X_1), \]

and

\[(Z, Y_1) \approx_{2^{-\Omega(n)}} (U, Y_1). \]

If $X_1 \neq X'_1$ or $Y_1 \neq Y'_1$ then we have $V_1 \neq V'_1$. Now consider the case where $X_1 \neq X'_1$ and $Y_1 \neq Y'_1$. In this case we have $Z = Z'$ and either $X_2 \neq X'_2$ or $Y_2 \neq Y'_2$. Without loss of generality assume that $X_2 \neq X'_2$. We can now first fix $(X_1, X'_1)$. Note that conditioned on this fixing, $Z=Z'$ is a deterministic function of $Y$, and thus independent of $(X_2, X'_2)$. The Reed-Solomon encoding of $\hat{X}_2$ and $\hat{X}_2'$ ensures that $\overline{X}_2$ and $\overline{X}_2'$ differ in at least $n-n_0+1 > 0.9 n$ symbols. Thus, with probability $1-2^{-\Omega(n)}-2^{-\Omega(r/\log n)}=1-2^{-\Omega(n/\log n)}$ over $Z$, we have that $\tilde{X}_2 \neq \tilde{X}_2'$. Therefore, altogether with probability $1-2^{-\Omega(n/\log n)}$ over the fixing of $(X_1, Y_1, \tilde{X}_2, \tilde{Y}_2)$ and $(X'_1, Y'_1, \tilde{X}_2', \tilde{Y}_2')$ we have that $V_1 \neq V'_1$.

We now fix $(X_1, Y_1, \tilde{X}_2, \tilde{Y}_2)$ and $(X'_1, Y'_1, \tilde{X}_2', \tilde{Y}_2')$. Note that conditioned on this fixing, $X$ and $Y$ are independent. Moreover, the average conditional min-entropy of both $X_3$ and $Y_3$ is at least $15 \tau n-\tau n-2 \tau n-3 \tau n=9 \tau n$. Thus by Theorem~\ref{thm:ip}, we have that

\[(W, X_3) \approx_{2^{-\Omega(n)}} (U, X_3). \]

We ignore the error for now since this only adds $2^{-\Omega(n)}$ to the final error. We now fix $(X_3, X'_3)$. Note that conditioned on this fixing, $(W, W')$ is a deterministic function of $(Y, Y')$, and thus independent of $(X, X')$. Further, the average conditional min-entropy of $X_4$ is at least $40 \tau n-\tau n-2 (15 \tau n+\tau n)-3 \tau n=4 \tau n$. Thus by Lemma~\ref{altext} we have that for any $0 \leq j \leq L-1$,

\[(W, W', \{R_1, R'_1, \cdots, R_j, R'_j\}, R_{j+1}) \approx_{\e'} (W, W', \{R_1, R'_1, \cdots, R_j, R'_j\}, U),\]

where $\e'=O(L 2^{-\Omega(n/\log n)})=2^{-\Omega(n/\log n)}$. Since conditioned on the fixing of $(W, W')$, the random variables $\{R_i, R'_i\}$ are deterministic functions of $(X, X')$ and independent of $(Y, Y')$, we also have that 

\[(Y_3, Y'_3, \{R_1, R'_1, \cdots, R_j, R'_j\}, R_{j+1}) \approx_{\e'} (Y_3, Y'_3, \{R_1, R'_1, \cdots, R_j, R'_j\}, U).\]

We now further fix $(Y_3, Y'_3)$. Note that now we have fixed $(X_1, Y_1, \tilde{X}_2, \tilde{Y}_2, X_3, Y_3)$ and $(X'_1, Y'_1, \tilde{X}_2', \tilde{Y}_2', X'_3, Y'_3)$. Ignoring the error for now let's assume that $V_1 \neq V'_1$ (note that $(V_1, V'_1)$ are now fixed) and for any $0 \leq j \leq L-1$,

\[(\{R_1, R'_1, \cdots, R_j, R'_j\}, R_{j+1}) = (\{R_1, R'_1, \cdots, R_j, R'_j\}, U).\]

Let $j$ be the index when the protocol executes step 8. We know that $j \leq L$ since in each step the length of the string $V_i$ goes from $n_i$ to $O(\log n_i)$. We have the following observation. For any $1 \leq i \leq j$, we have that $V_i$ is a deterministic function of $(R_1, \cdots, R_{i-1})$; similarly, $V'_i$ is a deterministic function of $(R'_1, \cdots, R'_{i-1})$. Next, we have the following claim.

\BCM
For any $1 \leq i < j$, suppose that conditioned on the fixing of $(R_1, \cdots, R_{i-1}), (R'_1, \cdots, R'_{i-1})$ we have $V_i \neq V'_i$, then with probability $1-2^{-\Omega(\log n_i)}$ over the further fixing of $(R_i, R'_i)$, we have $V_{i+1} \neq V'_{i+1}$. Suppose that conditioned on the fixing of $(R_1, \cdots, R_{j-1}), (R'_1, \cdots, R'_{j-1})$ we have $V_j \neq V'_j$, then with probability $1-2^{-\Omega(L)}$ over the further fixing of $(R_j, R'_j)$, we have $\tilde{\alpha} \neq \tilde{\alpha}'$.
\ECM

\begin{proof}[Proof of the claim.] Suppose that conditioned on the fixing of $(R_1, \cdots, R_{i-1}), (R'_1, \cdots, R'_{i-1})$ we have $V_i \neq V'_i$. Note that now $(V_i, V'_i)$ is also fixed. We know that $R_i$ is still uniform. Again, we have two cases. First, if $R_i \neq R'_i$, then we definitely have $V_{i+1} \neq V'_{i+1}$. Otherwise, we have $R_i = R'_i$. The encoding of $V_i$ and $V'_i$ ensures that at least a constant fraction of bits in $\tilde{V_i}$ and $\tilde{V_i}'$ are different. Thus by Theorem~\ref{thm:sampler} with probability $1-2^{-\Omega(\log n_i)}$ over the further fixing of $(R_i, R'_i)$, we have that $\overline{V_i} \neq \overline{V_i}'$ and thus $V_{i+1} \neq V'_{i+1}$.

For the case of $i=j$, the argument is the same, except now we are sampling $L-O(\log n_j)$ bits, and the probability that $\overline{V_i} \neq \overline{V_i}'$ is $2^{-\Omega(L -O(\log n_j))}=2^{-\Omega(L)}$ since $L \geq c \log n_j$.
\end{proof}

Now we are basically done. Since we start with $V_1 \neq V'_1$, at the end the probability that $\tilde{\alpha} \neq \tilde{\alpha}'$ is at least 

\[ \Pi_{i=1}^{j-1} (1-2^{-\Omega(\log n_i)}) \cdot (1-2^{-\Omega(L)}).\]

Note that for any $1 \leq i < j$ we have $n_{i+1}=O(\log n_i)$, so $2^{-\Omega(\log n_i)} \leq 2^{-\Omega(\log n_i)} / 2$. Thus the terms $2^{-\Omega(\log n_i)}$ form at least a geometric expression and hence this probability is at least $1-O(2^{-\Omega(L)})=1-2^{-\Omega(L)}$. Adding back all the errors, and noticing that $C \leq L \leq \frac{\mu n}{\log n}$ for some properly chosen constants $C$ and $\mu$, the final error is still $1-2^{-\Omega(L)}$. Moreover, since the size of each random variable in $(X_1, Y_1, \tilde{X}_2, \tilde{Y}_2, X_3, Y_3, X_4)$ is at most $O(\tau n)$, conditioned on the fixing of $(X_1, Y_1, \tilde{X}_2, \tilde{Y}_2, X_3, Y_3, X_4)$ and the tampered versions $(X'_1, Y'_1, \tilde{X}_2', \tilde{Y}_2', X'_3, Y'_3, X'_4)$, the average conditional min-entropy of both $X$ and $Y$ is $(1-O(\tau)) n$.
\end{proof}

We now use the above advice generator to give a new construction of non-malleable two-source extractors. Let $(X, Y)$ be two independent $(n, (1-\gamma)n)$ sources with $\gamma \leq \tau$ where $\tau$ is the constant in Lemma~\ref{lem:newadv}.
  
\begin{itemize}

\item Let $\adg$ be the advice generator from Lemma~\ref{lem:newadv} for some error $\e_1$.

\item Let $\acb$ be the correlation breaker with advice from Lemma~\ref{lem:advcb2} with error some $\e_2$, using the merger from Lemma~\ref{lem:nnipm2}. 

\item Let $\iext$ be the invertible linear seeded extractor form Theorem~\ref{thm:iext}.
\end{itemize}

\begin{enumerate}

\item Compute $\tilde{\alpha}=\adg(X, Y)$. 

\item Consider the unused part of $X$. Divide it into $(X_5, X_6, X_7)$ where $X_5, X_6$ has length $\alpha n, \beta n$ for some constants $\beta> \alpha>\gamma$, and $X_7$ is the rest of $X$ with length at least $n/2$. Similarly, divide the unused part of $Y$ into $(Y_5, Y_6, Y_7)$ where $Y_5, Y_6$ has length $\alpha n, \beta n$ and $Y_7$ is the rest of $Y$ with length at least $n/2$ (this can be ensured by choosing $\alpha, \beta, \gamma$ to be small enough).

\item Compute $V=\acb(X_5, Y_5, \tilde{\alpha})$ which outputs $d=O(\log(n/\e_2))$ bits.

\item Finally compute $W=\iext(Y_6, V)$ which outputs $\Omega(n)$ bits.
\end{enumerate}

We need the following proposition.

\begin{proposition}\cite{CG14b}
Let $D$ and $D'$ be distributions over the same finite space $\Omega$, and suppose they are $\e$-close to each other. Let $E \subseteq \Omega$ be any event such that $D(E) = p$. Then, the conditional distributions $D|E$ and
$D'|E$ are $(\e/p)$-close.
\end{proposition}

We now have the following theorem.

\BT \label{thm:newnm}
Assume that either the tampering function $f$ on $X$ or the tampering function $g$ on $Y$ has no fixed point. There exist a constant $C>1$ such that as long as $n \geq C \frac{\log \log (1/\e_1)}{\log \log \log (1/\e_1)} \log(n/\e_2)$, the above non-malleable two-source extractor gives a non-malleable code with error $\e_1+O(\log(1/\e_1) \sqrt{\e_2})$ and rate $\Omega(\log (1/\e_2)/n)$.
\ET

\begin{thmproof}
First note that by Lemma~\ref{lem:newadv}, conditioned on the fixing of $H=(X_1, Y_1, \tilde{X}_2, \tilde{Y}_2, X_3, Y_3, X_4)$ and the tampered versions $H'=(X'_1, Y'_1, \tilde{X}_2', \tilde{Y}_2', X'_3, Y'_3, X'_4)$, $X$ and $Y$ are independent, and the average conditional min-entropy of both $X$ and $Y$ is $(1-O(\gamma)) n$. If in addition we have that $\tilde{\alpha} \neq \tilde{\alpha}'$, then we will apply Lemma~\ref{lem:advcb2} and Lemma~\ref{lem:nnipm2}. Note that in order to set the error of the advice generator to be $\e_1$, we need to set the advice length to be $L=O(\log (1/\e_1))$ by Lemma~\ref{lem:newadv}. Thus in  Lemma~\ref{lem:advcb2} we need to merge $L=O(\log (1/\e_1))$ rows.

Again, as in Theorem~\ref{thm:tnmext}, we know that when we apply the correlation breaker to $X_5$  and $Y_5$, the entropy loss of both of them is $O(\gamma n)$. By choosing $\alpha, \beta, \gamma$ appropriately we can ensure that $X_5$  and $Y_5$ have sufficient entropy in them. We choose $a=2$ in Lemma~\ref{lem:nnipm2} and thus we obtain a correlation breaker with $m=O(\log(n/\e_2))$, $d_1=O(\log(n/\e_2))$ and $d_2=\log (n/\e_2)2^{O(\sqrt{\log t})}$ where $t$ is the parameter in Construction~\ref{con:advcb2} with $t \leq L$. Note that this also satisfies that $d_1 \geq 4m$ and $m \geq c \log (d_2/\e)$ as required by Lemma~\ref{lem:advcb2}.

Now we need to ensure that

\[(\alpha- O(\gamma))n \geq c  \frac{\log L}{\log t} \log(n/\e_2)+max\{ 8\frac{\log L}{\log t} d_1, 2t \cdot d'+4d_2\}+5 \ell+4 \log(1/\e_2),\]

where $d'=O(\log(n/\e_2))$. We choose $t =\frac{\log L}{\log \log L}$ and this gives us

\[n \geq C \frac{\log L}{\log \log L} \log(n/\e_2),\]

for some constant $C>1$. That is, we need 

\[n \geq C \frac{\log \log (1/\e_1)}{\log \log \log (1/\e_1)} \log(n/\e_2),\]

for some constants $C>1$. As long as this condition is satisfied, conditioned on the event that $\tilde{\alpha} \neq \tilde{\alpha}'$, we have that

\[(V, V', H, H', X, X') \approx_{O(L \e_2)} (U, V', H, H', X, X').\]

By choosing $\beta> \alpha$ appropriately, we can ensure that conditioned on the fixing of the previous random variables in the computation, $Y_6$ has entropy $\Omega(n)$ and $(V, V')$ is a deterministic function of $(X, X')$ and thus independent of $(Y, Y')$. Thus eventually we get

\[(W, W', H, H', X, X') \approx_{O(L \e_2)} (U, W', H, H', X, X').\]

However, note that our construction is a two-source extractor itself. Thus, regardless of whether $\tilde{\alpha} \neq \tilde{\alpha}'$, we have that

\[(W, H, H', X, X') \approx_{O(L \e_2)} (U, H, H', X, X').\]

We can cut the output length of the extractor to be $m=\Theta(\log (1/\e_2))$ such that for any $s$ in the support, we have $\Pr[U=s]=2^{-m}=\sqrt{\e_2}$. Thus we have for any $s$, 

\[(H, H', X, X'|W=s) \approx_{O(L \sqrt{\e_2})} (H, H', X, X'|U=s).\]

This means for any $s$, 

\[(H, H', X, X'|W=s) \approx_{O(L \sqrt{\e_2})} (H, H', X, X').\]

Let $A$ be the event that $\tilde{\alpha} \neq \tilde{\alpha}'$. Note that $\Pr[A] \geq 1-\e_1$. Since $A$ is determined by $(H, H')$, we have that for any $s$, $|\Pr[A|W=s]-\Pr[A] | \leq O(L \sqrt{\e_2})$. 

We now consider the probability distribution $(W'|W=s, A)$. This time we first condition on $A$. Note that conditioned on this event, we have

\[(W, W', H, H', X, X') \approx_{O(L \e_2)} (U, W', H, H', X, X').\]

Thus again here we have that for any $s$,

\[(W', H, H', X, X'|W=s) \approx_{O(L \sqrt{\e_2})} (W', H, H', X, X').\]

Therefore, we have for any $s$, 

\[(W'|W=s, A) \approx_{O(L \sqrt{\e_2})} (W'|A).\]

We can now bound the statistical distance between $(W'|W=s)$ and $W'$. We have

\begin{align*}
& \left |(W'|W=s) -W' \right |  \\= &\left |( \Pr[A|W=s] (W'|W=s, A)+\Pr[\bar{A}|W=s] (W'|W=s, \bar{A})) - (\Pr[A] (W'|A) +\Pr[\bar{A}](W'|\bar{A})) \right |
\\ \leq & \left |\Pr[A] ((W'|W=s, A)- W'|A) \right | + \left | (\Pr[A|W=s]-\Pr[A] )(W'|W=s, A) \right | 
\\ & + \left |\Pr[\bar{A}] ((W'|W=s, \bar{A})- (W'|\bar{A})) \right | +\left |(\Pr[\bar{A}|W=s]-\Pr[\bar{A}]) (W'|W=s, \bar{A}) \right |
\\ \leq & \left |\Pr[A] ((W'|W=s, A)- W'|A) \right | +\left |\Pr[\bar{A}] ((W'|W=s, \bar{A})- (W'|\bar{A})) \right |+O(L \sqrt{\e_2})
\\ \leq & \left |((W'|W=s, A)- W'|A) \right | + \Pr[\bar{A}]
\\ \leq & \e_1+O(L \sqrt{\e_2}).
\end{align*}

Note that the distribution of $W'$ is a fixed probability distribution which is independent of $s$. Thus the construction gives a non-malleable code with error $\e_1+O(L \sqrt{\e_2})=\e_1+O(\log(1/\e_1) \sqrt{\e_2})$, and the rate of the code is $\Omega(\log (1/\e_2)/n)$.
\end{thmproof}

We need to use the following simple inequality:

\begin{fact}\label{fa:inq}
For any $0< x \leq 1/3$, we have $1-3x \leq \frac{1-x}{1+x} < \frac{1+x}{1-x} \leq 1+3x$.
\end{fact}

We now have the following lemma, which gives a construction of non-malleable codes in the general case.

\BL \label{lem:newnm}
Assume $\twext: \bits^n \times \bits^n \to \bits^m$ satisfies the following conditions:

\begin{itemize}
\item It is a two-source extractor for entropy $n-\log(1/\e')$ with error $\e' \leq 2^{-(m+2)}$.
\item It is a non-malleable two-source extractor for entropy $n-\log(1/\e')$, which gives a non-malleable code in the two-split state model with error $\e$ when either the tampering function $f$ or the tampering function $g$ has no fixed point.
\end{itemize}

Then $\twext$ gives  non-malleable code in the two-split state model with error $\e+2^{m+4}\e'$.
\EL

\begin{proof}
Consider the tampering function $f: \bits^n \to \bits^n$ and $g: \bits^n \to \bits^n$. Let $X$ and $Y$ be two independent uniform distributions on $\bits^n$, let $p_0=\Pr[f(X)=X]$, $q_0=\Pr[g(Y)=Y]$, $p_1=\Pr[f(X)\neq X]=1-p_0$ and $q_1=\Pr[g(Y)\neq Y]=1-q_0$. Let the subsource of $X$ conditioned on $f(X)=X$ be $X_0$, and the subsource of $X$ conditioned on $f(X) \neq X$ be $X_1$. Thus $X=p_0 X_0+p_1X_1$. Similarly, we can define the subsources $Y_0, Y_1$ of $Y$ such that $Y=q_0 Y_0+q_1Y_1$.

Consider the pairs of subsources $(X_0, Y_0)$, $(X_0, Y_1)$, $(X_1, Y_0)$, and $(X_1, Y_1)$, which have probability mass $p_0 q_0$, $p_0q_1$, $p_1q_0$ and $p_1 q_1$ respectively. Note that we have

\[(X, Y)=p_0 q_0 (X_0, Y_0) +p_0 q_1 (X_0, Y_1)+p_1 q_0 (X_1, Y_0)+p_1 q_1 (X_1, Y_1).\]

Let $W=\twext(X, Y)$. Consider any $s \in \bits^m$ and the uniform distribution on the pre-image of $W=s$ in $(X, Y)$, call it $Z_s$. For any $i, j \in \bits$, let the subsource $Z_{ij}$ stand for the uniform distribution on the pre-image of $W=s$ in $(X_i, Y_j)$. Further let $r_{ij}=\Pr[\twext(X_i, Y_j)=s]$. Then we have 

\[Z_s=\frac{\sum_{i, j} p_i q_j r_{ij} Z_{ij}}{\sum_{i, j} p_i q_j r_{ij}}=\sum_{i, j} \alpha_{ij} Z_{ij},\]

where $\alpha_{ij}= \frac{p_i q_j r_{ij}}{\sum_{i, j} p_i q_j r_{ij}}$. 

We now have the following claim.

\BCM
For any $i, j \in \bits$, we have
\begin{itemize}
\item If either $p_i < \e'$ or $q_j < \e'$, then $\alpha_{ij} \leq 2^{m+1} \e'$.
\item Otherwise, $|\alpha_{ij}/(p_i q_j) - 1| \leq 2^{m+2} \e'$
\end{itemize}
\ECM

\begin{proof}[Proof of the claim.]

Note that $\sum_{i, j} p_i q_j r_{ij}=\Pr[W=s]$, and we have $\Pr[W=s] \geq  2^{-m} - \e' > 2^{-(m+1)}$. Thus if  either $p_i < \e'$ or $q_j < \e'$, we have 

\[\alpha_{ij}= \frac{p_i q_j r_{ij}}{\sum_{i, j} p_i q_j r_{ij}} < 2^{m+1} \e'.\]

Otherwise, both $p_i \geq \e'$ and $q_j \geq \e'$. This means that both $X_i$ and $Y_j$ have min-entropy at least $n-\log(1/\e')$. Therefore we have $|r_{ij}-2^{-m}| \leq \e'$. Note that $\alpha_{ij}/(p_i q_j) =r_{ij}/\Pr[W=s]$ and we also have .$|\Pr[W=s]-2^{-m}| \leq \e'$. Since $\e' \leq 2^{-(m+2)}$ by Fact~\ref{fa:inq} we have that 

\[|\alpha_{ij}/(p_i q_j) - 1| \leq 2^{m+2} \e'.\]
\end{proof}

 We now consider the distribution $\twext(T(Z_s))$, where for any distribution $Z$ on $\bits^n \times \bits^n$, $T(Z)$ stands for the distribution $(f(x), g(y))$ where $(x, y)$ is sampled from $Z$. Note that $\twext(Z_s)$ is fixed to $s$ and and $\twext(T(Z_s))$ is the distribution of the decoded message after tampering. We have that $T(Z_s)=\sum_{i, j} \alpha_{ij} T(Z_{ij})$ and $\twext(T(Z_s))= \sum_{i, j} \alpha_{ij} \twext(T(Z_{ij}))$. We will show that this distribution is close to the following distribution. For any $i, j \in \bits$ that are not both $0$, if either $p_i < \e'$ or $q_j < \e'$, we define the distribution $D_{ij}$ on $\bits^m$ to be a fixed constant (e.g., $\Pr[D_{ij}=0^m]=1$); otherwise since both $X_i$ and $Y_j$ have min-entropy at least $n-\log(1/\e')$, $\twext$ gives a non-malleable code and thus $\twext(T(Z_{ij})$ is $\e$-close to a distribution $D_{ij}$ independent of $s$. We let $D_{00}$ be the distribution obtained by the identity function, i.e., for any $s$, $D_{00}$ is fixed to be $I(s)=s$. We now claim that $\twext(T(Z_s))$ is close to the distribution $\sum_{i, j} p_i q_j D_{ij}$. We have
 
\[\left |\twext(T(Z_s))-\sum_{i, j} p_i q_j D_{ij} \right | =\left |\sum_{i, j} \alpha_{ij} \twext(T(Z_{ij})) -\sum_{i, j} p_i q_j D_{ij}\right | \leq \sum_{i, j}\left |\alpha_{ij} \twext(T(Z_{ij}))- p_i q_j D_{ij}\right |.\] 

For any $i, j \in \bits$, if either $p_i < \e'$ or $q_j < \e'$, we have the following bound.

\[\left |\alpha_{ij} \twext(T(Z_{ij}))- p_i q_j D_{ij}\right | \leq  |\alpha_{ij} \twext(T(Z_{ij}))|+|p_i q_j D_{ij}| \leq 2^{m+1} \e'+\e' < 2^{m+2} \e'.\]

Otherwise if $i, j$ are not both $0$ we have the following bound.

\[\left |\alpha_{ij} \twext(T(Z_{ij}))- p_i q_j D_{ij}\right | \leq p_i q_j |\twext(T(Z_{ij}))-D_{ij}|+|(\alpha_{ij}-p_i q_j) \twext(T(Z_{ij}))| \leq p_i q_j \e+2^{m+2} \e'.\]

For the case of $i=j=0$, we have that for any $(x, y) \in \Supp(Z_{00})$, $f(x)=x$ and $g(y)=y$. Thus $\twext(T(Z_{ij}))=s=D_{00}$ and we have

\[\left |\alpha_{ij} \twext(T(Z_{ij}))- p_i q_j D_{ij}\right | \leq p_i q_j |\twext(T(Z_{ij}))-D_{ij}|+|(\alpha_{ij}-p_i q_j) \twext(T(Z_{ij}))| \leq 2^{m+2} \e'.\]

Therefore altogether we have

\[\left |\twext(T(Z_s))-\sum_{i, j} p_i q_j D_{ij} \right | \leq \sum_{i, j } (p_i q_j \e+2^{m+2} \e') =\e+2^{m+4} \e'.\]

Since $\sum_{i, j} p_i q_j D_{ij}$ is obtained by $G(s)$ where $G$ is a fixed probability distribution on the identity function and constant functions (the distribution of $G$ only depends on $f$ and $g$, but not on $s$), this implies that we have a non-malleable code in the $2$ split-state model with error $\e+2^{m+4} \e'$.
\end{proof}

We now have the following theorem.

\BT \label{thm:newcode}
There are constants $0< \eta, \mu<1$ such that for any $n \in \N$ and $2^{-\frac{\mu n}{\log n}} \leq \e \leq \eta$ there exists an explicit non-malleable code in the $2$-split-state model with block length $2n$, rate  $\Omega(\frac{\log \log \log (1/\e)}{\log \log (1/\e)})$ and error $\e$.
\ET

\begin{thmproof}
We combine Theorem~\ref{thm:newnm} and Lemma~\ref{lem:newnm}. Note that in Theorem~\ref{thm:newnm}, the construction is itself a two-source extractor for entropy $(1-\gamma) n$ with error $O(\log(1/\e_1) \e_2)$. To apply Theorem~\ref{thm:newnm}, we just need to ensure that 

\[n \geq C \frac{\log \log (1/\e_1)}{\log \log \log (1/\e_1)} \log(n/\e_2)\]

for some constant $C>1$. We set $\e_1=\e/2$ and $\e_2=2^{-\Omega(\frac{\log \log \log (1/\e) n}{\log \log (1/\e)})}$. Note that 

\[C \frac{\log \log (1/\e_1)}{\log \log \log (1/\e_1)} \log(n/\e_2)=O( \frac{\log \log (1/\e)}{\log \log \log (1/\e)}\log n)+O( \frac{\log \log (1/\e)}{\log \log \log (1/\e)}\log (1/\e_2)).\]

Since $2^{-\frac{\mu n}{\log n}} \leq \e$ we have $\frac{\log \log (1/\e)}{\log \log \log (1/\e)}\log n=O(\frac{\log^2 n}{\log \log n})$. Thus we can set $\e_2=2^{-\Omega(\frac{\log \log \log (1/\e) n}{\log \log (1/\e)})}$ to satisfy the inequality. Now we apply Lemma~\ref{lem:newnm}. We can set $\e'=O(\log(1/\e_1) \e_2)$ since by Theorem~\ref{thm:newnm} the construction is both a two-source extractor and a non-malleable two-source extractor for entropy $(1-\gamma) n$, and as long as $\e \leq \eta$ for some appropriately chosen $\eta<1$ we have $\log(1/\e') \leq \gamma n$. Since in Theorem~\ref{thm:newnm} we set the output of the extractor to be $m=\Theta(\log (1/\e_2))$ such that $2^{-m}=\sqrt{\e_2}$, we have that $\e' \leq 2^{-(m+2)}$ and $2^{m+4}\e'=O(\log(1/\e_1) \sqrt{\e_2})$. Thus by Lemma~\ref{lem:newnm} the final error of the non-malleable code is 

\[\e_1+O(\log(1/\e_1) \sqrt{\e_2})+2^{m+4}\e'=\e/2+O(\log(1/\e)\sqrt{\e_2}).\]
Finally, notice that 

\[\sqrt{\e_2}=2^{-\Omega(\frac{\log \log \log (1/\e) n}{\log \log (1/\e)})} \leq \alpha \frac{\e}{\log (1/\e)}\]

for any arbitrary constant $\alpha>0$, since the latter is at least $\frac{1}{n}2^{-\frac{\mu n}{\log n}}$ and $\e_2$ is $2^{-\Omega(\frac{n \log \log n}{\log n})}$. Thus the final error of the non-malleable code is at most $\e/2+\e/2=\e$, while the rate of the code, by Theorem~\ref{thm:newnm}, is $\Omega(\log (1/\e_2)/n)=\Omega(\frac{\log \log \log (1/\e)}{\log \log (1/\e)})$.
\end{thmproof}

Next, we show how to achieve better error in the non-malleable two-source extractor and non-malleable codes. Recall that a bottleneck for error is the use of Reed-Solomon code in the construction. In order to get better error, we instead use a binary linear error correcting code and its generating matrix. It is easy to show using standard probabilistic argument that there exists a binary generating matrix that satisfies our requirements.

\BT
There exists constants $0< \alpha, \beta<1$ such that for any $n \in \N$ there exists an $ n \times m$ matrix over $\F_2$ with $n=\beta m$ which is the generating matrix of an asymptotically good code. Furthermore, Any sub-matrix formed by taking $\alpha n$ columns and the last $n/2$ rows has full column rank. In addition, for some $\e=2^{-O(n)}$, an $\e$-biased sample space over $nm$ bits generates such a matrix with probability $1-2^{-\Omega(n)}$.
\ET

\begin{thmproof}
We take an $\e$-biased sample space over $nm$ bits for some $\e=2^{-O(n)}$. First, consider the sum of the rows over any non-empty subset of the rows. The sum is an $m$-bit string such that any non-empty parity is $\e$-close to uniform. Thus by the XOR lemma it is $2^{m/2}\e$-close to uniform. We know a uniform $m$-bit string has weight $d=m/4$ with probability at least $1-2^{-\Omega(m)}$. Thus for this string the probability is at least $1-2^{-\Omega(m)}-2^{m/2}\e$. By a union bound the total failure probability is at most $2^n(2^{-\Omega(m)}+2^{m/2}\e)=2^{-\Omega(n)}$ by an appropriate choice of $\beta$ and $\e=2^{-O(n)}$.

Next, consider any sub-matrix formed by taking $\beta m$ columns and the last $n/2$ rows, if it's truly uniform, then the probability that it has full column rank is at least $1-\alpha n 2^{\alpha n-n/2} \geq 1-2^{-n/4}$ for $\alpha<1/5$. Now by a union bound the total failure probability is at most 

\[\binom{m}{\alpha n}(2^{-n/4}+\e) \leq \left (\frac{e m}{\alpha n} \right )^{\alpha n} 2^{-n/4+1}=\left (\frac{e} {\beta \alpha}\right )^{\alpha n}  2^{-n/4+1},\]

if we choose $\e < 2^{-n/4}$. Note that for a fixed $\beta$, the quantity $(\frac{e} {\beta \alpha})^{\alpha}$ goes to $1$ as $\alpha$ goes to $0$. Thus we can choose $\alpha$ small enough such that this failure probability is also $2^{-\Omega(n)}$. Therefore altogether the failure probability is $2^{-\Omega(n)}$.
\end{thmproof}

Note that an $\e$-biased sample space over $nm$ bits can be generated using $O(\log (nm/\e))=O(n)$ bits if $\e=2^{-O(n)}$. Now for any length $n \in \N$, we can compute the generating matrix (either using an $\e$-biased sample space or compute it deterministically in $2^{O(n)}$ time) once in the pre-processing step, and when we do encoding and decoding of the non-malleable code, all computation can be done in polynomial time. 

Combining Theorem~\ref{connection} and Theorem~\ref{thm:tnmext}, we immediately obtain the following theorem.

\BT \label{thm:nmcode}
For any $n \in \N$ there exists a non-malleable code with efficient encoder/decoder in the $2$-split-state model with block length $2n$, rate  $\Omega(\log \log n/\log n)$ and error $=2^{-\Omega(n \log \log n/\log n)}$.
\ET